%% file: main.tex
\documentclass[
aps,
prl, 
reprint,
superscriptaddress,
nofootinbib,
nobibnotes,
]{revtex4-2}

\usepackage{graphicx}
\usepackage[usenames,dvipsnames,table,xcdraw]{xcolor}
\usepackage{xspace}
\usepackage{amsmath,amsfonts,amssymb,amsthm}
\usepackage{mathtools}
\usepackage[varg]{txfonts} % Changes both the text and math font to times
\usepackage{bm}
\usepackage[utf8]{inputenc}
\usepackage[normalem]{ulem}
\usepackage{makecell}
\usepackage{multirow}
\usepackage{booktabs}
\usepackage{float}
\usepackage{adjustbox}
\usepackage[caption=false]{subfig}
\usepackage{rotating}
\usepackage{dcolumn}
\usepackage{enumitem}
\xdefinecolor{mylinkcolor}{rgb}{0,0,0.7}
\usepackage[
bookmarksnumbered,
bookmarksopen,
bookmarksopenlevel=2,
breaklinks=true,
colorlinks=true,
filecolor=mylinkcolor,
citecolor=mylinkcolor,
linkcolor=mylinkcolor,
urlcolor=mylinkcolor,
menucolor=mylinkcolor,
]{hyperref}

\usepackage{setspace}
\usepackage[hang]{footmisc} % 'hang' separates the number of the footnote from the paragraph
\renewcommand{\footnoterule}{%
  \kern -5pt  % This needs to be balanced with...
  \hrule width 1 \columnwidth
  \kern 5pt  % ... this value
}

\allowdisplaybreaks

\def\mF{\mathcal{F}}

\def\s{\hspace{0.5pt}}
\def\solarmass{\mathrm{M}_\odot}

\graphicspath{ {fig/} }

\makeatletter
\def\oldparagraph{
\@startsection
{paragraph}
{4}
{\parindent}
{\z@}
{-1sp}
{\normalfont\normalsize\itshape}
}
\makeatother
\renewcommand{\paragraph}[1]{\oldparagraph{\texorpdfstring{\textbf{#1}---}{#1}}}

\DeclareRobustCommand{\EndMatter}[1][sec:end_matter]{\hyperref[#1]{\emph{End Matter}}\xspace}
\DeclareRobustCommand{\SuppMat}[1][sec:supplemental]{\hyperref[#1]{\emph{Supplemental Material}}\xspace}

\newcommand{\AEI}{\affiliation{Max Planck Institute for Gravitational Physics (Albert Einstein Institute), Am M\"uhlenberg 1, Potsdam 14476, Germany}}
\newcommand{\Maryland}{\affiliation{Department of Physics, University of Maryland, College Park, MD 20742, USA}}

\newcommand{\Cornell}{\affiliation{Cornell Center for Astrophysics and Planetary Science, Cornell University, Ithaca, New York 14853, USA}}
\newcommand{\Caltech}{\affiliation{Theoretical Astrophysics 350-17, California Institute of Technology, Pasadena, California 91125, USA}}

\newcommand{\UoN}{\affiliation{Nottingham Centre of Gravity \& School of Mathematical Sciences, University of Nottingham, University Park, Nottingham NG7 2RD, United Kingdom}}
\newcommand{\UIB}{\affiliation{Departament de F\'isica, Universitat de les Illes Balears, IAC3 – IEEC, Crta. Valldemossa km 7.5, E-07122 Palma, Spain}}

\begin{document}

\title{
Enabling gravitational-wave astronomy with spin-precessing black holes on generic orbits
}

\author{Aldo Gamboa}
\email{aldo.gamboa@aei.mpg.de}
\AEI

\author{Lorenzo Pompili}
\UoN
\AEI

\author{Alessandra Buonanno}
\AEI
\Maryland

\author{Luca Sebastiani}
\AEI

\author{Raffi Enficiaud}
\AEI

\author{Michael Boyle}
\Cornell

\author{Lawrence E. Kidder}
\Cornell

\author{Harald P. Pfeiffer}
\AEI

\author{Antoni Ramos-Buades}
\UIB
\AEI

\author{Mark A. Scheel}
\Caltech

%%%%%%%%%%%%%%%%%%%%%%%%%%%
%%%%%%%%%%%%%%%%%%%%%%%%%%%
%%%%%%%%%%%%%%%%%%%%%%%%%%%

\begin{abstract}
Binary black holes (BBHs) formed in dense stellar environments or in hierarchical triples can coalesce on eccentric orbits and carry spins of arbitrary orientation, leaving distinctive imprints on their gravitational-wave (GW) emission.
We present \texttt{SEOBNRv6EPHM}: the first generic-orbit, spin-precessing model in the effective-one-body \texttt{SEOBNR} family, whose waveforms have underpinned LIGO--Virgo GW analyses since 2011.
The model describes the dynamics and multipolar GW signal of generic BBHs, covering the inspiral-merger-ringdown of coalescing binaries and extending to dynamical captures and scattering encounters.
We perform the first systematic accuracy assessment of a generic-orbit model against numerical relativity (NR) waveforms of spin-precessing BBHs, using 1437 quasi-circular (QC) and 87 eccentric simulations:
median waveform mismatches remain below $1 \%$, matching the accuracy of the QC model \texttt{SEOBNRv5PHM}, and improving on the state-of-the-art generic-orbit model \texttt{TEOBResumS-Dal\'i} by a median factor of $ 4 $.
The model also reproduces the non-perturbative phenomenology observed in NR simulations of generic-spin BBH scattering.
It is $\sim 2$--$3$ times faster than \texttt{SEOBNRv5PHM} in the QC limit, and up to an order of magnitude faster than \texttt{TEOBResumS-Dal\'i}, bringing eccentric inference to the cost of current QC analyses.
As a proof of principle, we analyze eleven GW events and focus on GW200129, strengthening its evidence for eccentricity---a result supported by injection-recovery studies with synthetic NR signals of eccentric, spin-precessing BBHs.
\texttt{SEOBNRv6EPHM} thus enables, for the first time, accurate and efficient GW analyses that jointly account for eccentricity and spin precession.
\end{abstract}

\date{\today}

\maketitle

%\tableofcontents

%%%%%%%%%%%%%%%%%%%%%%%%%%%
%%%%%%%%%%%%%%%%%%%%%%%%%%%
%%%%%%%%%%%%%%%%%%%%%%%%%%%

\paragraph{Introduction}

As the gravitational-wave (GW) catalog of compact binary coalescences grows~\cite{LIGOScientific:2016aoc,LIGOScientific:2018mvr,LIGOScientific:2020ibl,LIGOScientific:2021usb,LIGOScientific:2021djp,LIGOScientific:2025slb,LIGOScientific:2026wfs}, detecting \emph{orbital eccentricity} is ever more likely.
Compact binaries formed in galactic fields~\cite{Bethe:1998bn,Belczynski:2001uc,Mandel:2018hfr} coalesce on quasi-circular (QC) orbits, as GW emission damps their eccentricity long before merger~\cite{Peters:1963ux,Peters:1964zz}.
Nonetheless, a non-negligible fraction of systems forms close to merger via dynamical interactions~\cite{Rodriguez:2015oxa,Zevin:2018kzq,Takatsy:2018euo} or is influenced by a tertiary companion~\cite{Zeipel:1910, kozai1962secular, lidov1962evolution,Naoz:2016cjb}, retaining detectable eccentricity.
The same environments also produce spins misaligned with the orbital angular momentum, leading to spin-induced orbital precession (\emph{spin precession})~\cite{Apostolatos:1994mx}.

Ignoring eccentricity leads to missed GW signals~\cite{Brown:2009ng,Huerta:2013qb,Gadre:2024ndy} and to parameter-estimation (PE) biases in masses and spins~\cite{Favata:2013rwa,OShea:2021faf,Cho:2022cdy,GilChoi:2022waq,Romero-Shaw:2022fbf,Xu:2022zza,Divyajyoti:2023rht,Divyajyoti:2025cwq,Tibrewal:2026jci,RoyChowdhury:2026xgb,Chandra:2026voe}, in lensing evidence~\cite{Ezquiaga:2020gdt,Mishra:2025dpa}, in center-of-mass acceleration effects~\cite{Tiwari:2025aec,Pathak:2026cik,Roy:2026mco,Roy:2026duh,Pompili:2026tdf}, and in tests of general relativity~\cite{Saini:2022igm,Saini:2023rto,Narayan:2023vhm,Gupta:2024gun,Shaikh:2024wyn,Bhat:2022amc,Bhat:2024hyb,Chiaramello:2025bhi,Saini:2026jea}.
Conversely, eccentricity sharpens parameter measurements~\cite{Mikoczi:2012qy,Gondan:2017hbp,Gondan:2018khr,Tibrewal:2026shx}, enables early GW detection~\cite{Yang:2023zxk,Yang:2024vfy,Sinha:2025vmc}, enhances environmental effects~\cite{Takatsy:2025bfk,Zwick:2025qzv}, and, like misaligned spins~\cite{Biscoveanu:2026ikx}, is a \emph{smoking gun} of dynamical formation~\cite{Breivik:2016ddj,Nishizawa:2016eza,Zevin:2021rtf,Sedda:2026xqr,Rozner:2026jtj}.

A handful of GW events show potential evidence for eccentricity~\cite{Romero-Shaw:2020thy,Gayathri:2020coq,Romero-Shaw:2022xko,Iglesias:2022xfc,Gupte:2024jfe,Planas:2025jny,Morras:2025xfu,Xu:2025ajj,Planas:2025plq,Kacanja:2025kpr,Clarke:2026cuw,Pompili:2026yxq}, with astrophysical implications discussed in Refs.~\cite{Zeeshan:2024ovp,Romero-Shaw:2025otx,Romero-Shaw:2025vbc,Morras:2026mrv,Gupte:2026whi,Zeeshan:2026pga,Malagon:2026uev,Salvarese:2026klm,Zeeshan:2026vsi}.
However, these claims remain tentative due to noise artifacts (\emph{glitches}), high false-alarm rates, an incomplete understanding of eccentric PE, waveform-model inaccuracies, and the lack of joint eccentric, spin-precessing PE analyses.

Consolidating this evidence requires accurate waveform models that include both eccentricity and spin-precession effects.
Numerical relativity (NR) provides the most accurate waveforms for eccentric binaries~\cite{Hinder:2008kv,Lewis:2016lgx,Hinder:2017sxy,Huerta:2019oxn,Habib:2019cui,Ramos-Buades:2019uvh,Ramos-Buades:2022lgf,Ficarra:2026cej,Ferguson:2023vta,Trenado:2025ccf,Nee:2025zdy,Scheel:2025jct}, though at high computational cost.
Eccentric NR simulations remain scarce, however, and eccentric NR surrogate models exist only for nonspinning binary black holes (BBHs)~\cite{Islam:2021mha,Nee:2025nmh,Ravichandran:2026iec}.
Hence, eccentric GW analyses rely on semi-analytical waveform models.
Inspiral-merger-ringdown (IMR) models based on post-Newtonian (PN) results exist for eccentric, aligned-spin BBHs~\cite{Ramos-Buades:2021adz,Gamboa:2024hli,Gamboa:2026jht,Nagar:2024dzj,Nagar:2024oyk,Planas:2025feq,Ramos-Buades:2026kbq,Paul:2024ujx,Maurya:2025shc}.
Models for eccentric, spin-precessing BBHs have focused on the inspiral regime~\cite{Klein:2018ybm,Klein:2021jtd,Arredondo:2024nsl,Morras:2025nlp,Morras:2026fho}, and only recently on the full IMR process~\cite{Liu:2023ldr,Gamba:2024cvy,Albanesi:2025txj} using the effective-one-body (EOB) formalism~\cite{Buonanno:1998gg,Buonanno:2000ef}, which naturally incorporates eccentric effects~\cite{Bini:2012ji,Hinderer:2017jcs,Chiaramello:2020ehz,Khalil:2021txt,Gamboa:2024imd}.
These developments have enabled first IMR eccentric, spin-precessing PE analyses~\cite{Gamba:2025qfg,Jan:2025fps,Jan:2025zcm,Chandra:2025jfc} with the \texttt{TEOBResumS-Dal\'i} model~\cite{Gamba:2024cvy,Albanesi:2025txj}.
However, the computational cost of such inference has so far restricted these analyses to only four GW events, and the accuracy of generic-orbit models has yet to be systematically assessed against eccentric, spin-precessing NR simulations.

\begin{figure*}
\hspace{-8pt}
\includegraphics[width=\linewidth]{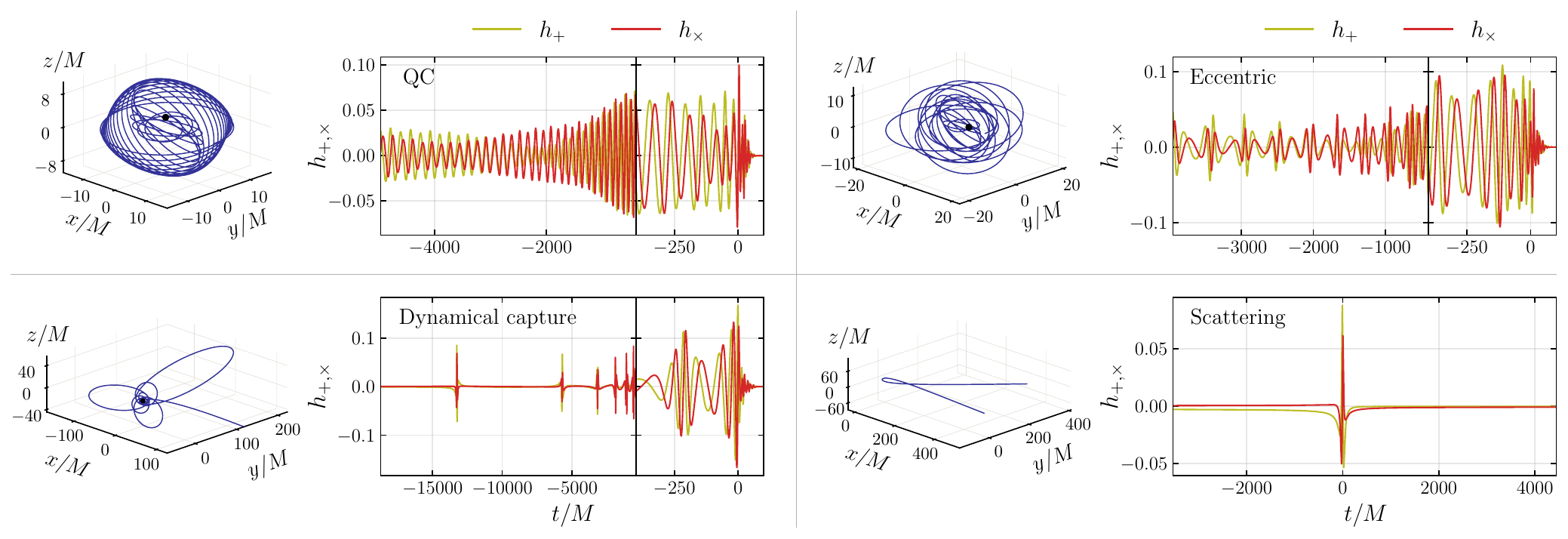}
\vspace{-5pt}
\caption{
EOB trajectories (left side of each panel) and the corresponding GW polarizations $ h_+ $ and $ h_\times $ at inclination $ \iota = \pi/3 $ (right side), for a spin-precessing BBH with mass ratio $ q = 3 $ and spins $ \bm{\chi}_1 = 0.7 \s \bm{\hat x} + 0.15 \s \bm{\hat z} $, $ \bm{\chi}_2 = 0.3 \s \bm{\hat x} + 0.3 \s \bm{\hat y} $, in four orbital configurations:
\emph{QC} with starting orbital frequency $ M \Omega = 0.017 $ (top left);
\emph{eccentric} with eccentricity $ e = 0.4 $, relativistic anomaly $ \zeta = 0 $, and orbit-averaged starting orbital frequency $ \langle M \Omega \rangle = 0.015 $ (top right);
\emph{dynamical capture} with energy $ E/M = 1.000001 $ and projected total angular momentum $ J_{\perp}/M^2 = 0.9 $ (bottom left);
and \emph{scattering} with $ E/M = 1.001 $ and $ J_{\perp}/M^2 = 0.95 $ (bottom right).
The values of $ E $ and $ J_{\perp} $ are specified at a separation of $ r/M = 500$.
For the merging configurations, the rightmost panels zoom into the merger-ringdown.
}
\label{fig:trajectories_waveforms}
\end{figure*}

Here, we introduce \texttt{SEOBNRv6EPHM},\footnote{
In the generic name \texttt{SEOBNRvnEPHM}, \texttt{S} stands for spinning, \texttt{NR} for the calibration to NR simulations, \texttt{vn} for the version of the model, \texttt{E} for eccentricity, \texttt{P} for spin precession, and \texttt{HM} for higher-order modes (i.e., multipoles beyond the dominant quadrupole).
}
an accurate and efficient model for generic-orbit, spin-precessing BBHs (see Fig.~\ref{fig:trajectories_waveforms}).
It is the first generic-orbit member of the \texttt{SEOBNR} family~\cite{Buonanno:2007pf,Taracchini:2012ig,Taracchini:2013rva,Pan:2013rra,Bohe:2016gbl,Cotesta:2018fcv,Ossokine:2020kjp,Pompiliv5,RamosBuadesv5,Estelles:2025zah,Haberland:2026xvj,FooInPrep} (which has provided IMR waveform templates for GW analyses since 2011~\cite{LIGOScientific:2011hqo}, up to the latest LIGO--Virgo~\cite{LIGOScientific:2014pky,VIRGO:2014yos} and KAGRA~\cite{KAGRA:2020tym} (LVK) catalogs~\cite{LIGOScientific:2026wfs}) and it reduces to \texttt{SEOBNRv6EHM}~\cite{Gamboa:2026jht} in the aligned-spin limit.
GWs are obtained by \emph{twisting-up} waveforms computed in a \emph{co-precessing frame}, in which the dynamics and radiation resemble those of an aligned-spin binary~\cite{Buonanno:2002fy,Schmidt:2010it,Boyle:2011gg,OShaughnessy:2011pmr,Schmidt:2012rh}, even for eccentric systems~\cite{Gamba:2024cvy,Thomas:2026qrg}.
Both the dynamics and the waveform are built from formulas that resum analytical PN information, with parameters calibrated against QC, aligned-spin NR and test-body-limit waveforms~\cite{Gamboa:2026jht}. 
For coalescing binaries, we smoothly attach a QC merger-ringdown portion informed by spin-precessing phenomenological results.

We work with geometric units $G \!=\! c \!= \! 1$ and, for a binary with source-frame component masses $m_{1,2}$ and spins $\bm{S}_{1,2}$, we define $ q = m_1/m_2 \geq 1$, $ M = m_1 + m_2 $, $ \mu = m_1 m_2 / M $, $ \nu = \mu / M $, $ \bm \chi_i = \bm a_i / m_i = \bm S_i / m_i^2 $, and $ \bm a _\pm = \bm a_1  \pm \bm a_2 $.

%%%%%%%%%%%%%%%%%%%%%%%%%%%
%%%%%%%%%%%%%%%%%%%%%%%%%%%
%%%%%%%%%%%%%%%%%%%%%%%%%%%

\paragraph{Generic-orbit EOB dynamics}

We evolve generic BBHs using planar EOB equations of motion coupled to PN-expanded equations for the spins and angular momentum:
\begin{subequations}
\label{eq:EOM}
\begin{align}
\dot{r}
&= \xi \frac{\partial H_\text{EOB}^\text{pprec}}{\partial p_{r_*}},
& \Omega \equiv \dot\phi
&= \frac{\partial H_\text{EOB}^\text{pprec}}{\partial p_\phi},
\label{eq:phidot_2}
\\
\dot{p}_{r_*}
&= -\xi \frac{\partial H_\text{EOB}^\text{pprec}}{\partial r} + \xi\,\mF_r,
\label{eq:dot_prstar}
& \dot{p}_\phi
&= \mF_\phi,
\end{align}
\vspace{-20pt}
\begin{align}
\dot{\bm{S}}_{i}
&= \bm{\Omega}_{S_{\!i}}^{\text{qc}}(\bm{S}_{1}, \bm{S}_{2}, \bm{l}_{\text{N}}, \Omega) \times \bm{S}_{i},
\label{eq:sdot}
\\[2pt]
\dot{\bm{l}}_{\text{N}}
&= \dot{\bm l}_{ \text{N}} ^{ \text{\s qc}}\s(\bm{S}_{1}, \bm{S}_{2}, \bm{l}_{\text{N}}, \Omega),
\label{eq:LNdot}
\\[3pt]
\bm{L}
&= \bm{L}^{\text{qc}}\s(\bm{S}_{1}, \bm{S}_{2}, \bm{l}_{\text{N}}, r).
\label{eq:L}
\end{align}
\end{subequations}
These are thirteen ordinary differential equations for the center-of-mass EOB polar coordinates $ (r, \phi, p_{r_*}, p_\phi) $ in the co-precessing frame, the spin vectors $ \bm{S}_{ 1} $ and $ \bm{S}_{ 2} $, and the Newtonian orbital angular momentum vector $ \bm L _{ \text{N}} = \mu \, \bm r \times \bm v $ (with unit direction $ \bm l _{ \text{N}} $), supplemented with an algebraic equation for the relativistic orbital angular momentum vector $ \bm L  = \mu \, \bm r \times \bm p $ (with unit direction $ \bm l $).
The momentum $ p_{r_*} \equiv p_r \, \xi(r)$ is conjugate to the tortoise coordinate $r_*$~\cite{Damour:2007xr,Pan:2009wj} with $\xi$ given in Ref.~\cite{Khalilv5}.

Initial conditions for the polar coordinates are set as in \texttt{SEOBNRv6EHM} (see Sec.~IV in Ref.~\cite{Gamboa:2026jht}) in a frame whose $ z $-axis is initially perpendicular to the orbital plane.
For bound binaries, we parametrize the orbit with the orbit-averaged orbital frequency $ \langle M \Omega \rangle $, eccentricity $ e $, and relativistic anomaly $ \zeta $, defined within the Keplerian parametrization of the orbit $ r/M = p/(1 + e \cos\zeta) $~\cite{darwin1959gravity,Gamboa:2024imd}, with $ p $ the semi-latus rectum.
For generic orbits, we use the separation $ r $, energy $ E/M = H _{ \text{EOB}} ^{ \text{pprec}}/M $, and the projection $ J _{ \perp} / M^2 \equiv \bm{l}_{ \text{N}} \cdot \bm{J}/M^2 $, where $\bm{J} = \bm{L} + \bm{S}_1+\bm{S}_2$ is the total angular momentum.

The conservative trajectory follows from the EOB Hamiltonian $ H_\text{EOB} = \nolinebreak M \sqrt{1 + 2 \nu \left( H_\text{eff}/\mu - 1 \right)} $, with $ H_\text{eff} $ the Hamiltonian of a test body of mass $ \mu $ moving in a $ \nu $-deformed Kerr spacetime.
We use a partial-precessing Hamiltonian, $ H _{ \text{EOB}} ^{ \text{pprec}} $, with orbit-averaged in-plane spin contributions for circular orbits~\cite{Khalilv5}. It is given in Appendix~A of Ref.~\cite{RamosBuadesv5}, and depends on $ (r, p_{r_*}, p_\phi) $, the spin magnitudes $ a_{\pm} $ and projections $ (\bm{l}_{ \text{N}} \cdot \bm{a}_{\pm} , \bm{l} \cdot \bm{a}_{\pm}) $, and on two calibration parameters $ a_6 (\nu) $ and $ \hat d _{ \text{SO}} (\nu, \bm{l}_{ \text{N}} \cdot \bm{a}_{\pm}) $ tuned to QC, aligned-spin NR simulations~\cite{Gamboa:2026jht}.

The dissipative dynamics is described by the radiation-reaction (RR) force components $ \mF_\phi = \nolinebreak \mF_\phi^\text{modes}\, \mF_\phi^\text{ecc} $ and $\mF_r = \nolinebreak p_{r}/p_\phi \ \mF_r^\text{modes} \,\mF_r^\text{ecc} $, which are consistent with the eccentric energy and angular momentum GW fluxes~\cite{Gamboa:2024imd,Gamboa:2026jht}.
The factors $ \mF_{\phi, \,\s r}^\text{modes} $ contain the leading-order terms for generic orbits augmented with high-PN-order contributions for QC orbits, and the factors $ \mF_{\phi, \,\s r}^\text{ecc} $ contain resummed eccentricity corrections at 1PN order.
These functions are given in Eqs.~(11) of Ref.~\cite{Gamboa:2026jht}, and depend on $ \big(\Omega, r, \dot r, \dot p_{r_*, \s\text{cons}} \equiv - \xi \, \partial H_\text{EOB}^\text{pprec}/\partial r \big) $ to ensure an accurate QC-orbit limit.
The QC aligned-spin contributions are updated with the projections $ \bm{l}_{ \text{N}} \cdot \bm{\chi}_{i} $. 

We use PN-expanded QC equations $ \big \{ \bm{\Omega}_{S_{\! i}}^{ \text{qc}}, \dot{\bm l}_{ \text{N}} ^{ \text{\s qc}}, \bm{L} ^{ \text{qc}} \big \} $ to model spin-precession effects~\cite{Akcay:2020qrj,Gamba:2021ydi}, as in \texttt{SEOBNRv5PHM}~\cite{RamosBuadesv5} (see the \SuppMat[sm:precession]).
We expect these equations to remain accurate for generic orbits when informed by an eccentric dynamics~\cite{Gamba:2024cvy}.
\texttt{TEOBResumS-Dal\'i} does so by feeding them with the frequency of a constant-spin eccentric planar dynamics.
Here, we instead solve Eqs.~\eqref{eq:EOM} as a fully coupled system, as we expect strong periastron passages to have a significant effect on the spin dynamics.

%%%%%%%%%%%%%%%%%%%%%%%%%%%
%%%%%%%%%%%%%%%%%%%%%%%%%%%
%%%%%%%%%%%%%%%%%%%%%%%%%%%

\paragraph{Generic-orbit waveforms}

We obtain the observer-frame polarizations $ h_+ $ and $ h_\times $ by twisting up a co-precessing waveform, following closely the \texttt{SEOBNRv5PHM} procedure~\cite{RamosBuadesv5}.

We define three reference frames (see Fig.~1 in Ref.~\cite{RamosBuadesv5}):
1) the inertial frame of the observer (\emph{I} frame),
2) an inertial frame whose $z$-axis is aligned with the final angular momentum $ \bm{J}_{ \text{f}} $ (\emph{J} frame),
and 3) a non-inertial co-precessing frame (\emph{P} frame) whose $z$-axis follows $ \bm{l} _{ \text{N}} $ during the inspiral-plunge, and after merger follows a QC NR-informed phenomenological evolution~\cite{OShaughnessy:2012iol,RamosBuadesv5}.
The frames are connected by a constant quaternion $ q_{I \to J} $ and a time-dependent one, $ q_{J \to P}(t) $ (equivalent to time-varying Euler angles $ \{ \alpha, \beta, \gamma \} $)~\cite{RamosBuadesv5}, that satisfies the minimal rotation condition $ \dot{\gamma}=-\dot{\alpha}\cos\beta $ to reduce precessional effects in the \emph{P}-frame waveform~\cite{Boyle:2011gg}.

The GW polarizations are constructed using the inverse rotations $ q _{P \to J}(t) \equiv \bar q _{J \to P}(t) $ and $ q _{J \to I} \equiv \bar q _{I \to J} $, demanding that the \emph{P} and \emph{I} frames coincide at the starting time.
Rather than computing the polarizations from the \emph{I}-frame \emph{waveform modes} $ h _{ \ell m} ^{ I} $ (via $ h_{+} - i h_{\times} = \sum_{\ell, m} {}_{-2} Y_{\ell m}(\iota, \varphi)\,h^{I}_{\ell m} $), we instead efficiently compute them using $ h_{+} - i h_{\times} =
e^{2i\, \gamma_{P \rightarrow I}}\sum_{\ell, m} {}_{-2} Y_{\ell m}(\beta_{ P \rightarrow I}, \alpha_{ P \rightarrow I}) \, h^{ P}_{\ell m} $.
Here, $ _{-2}Y_{\ell,\s m}$ are the $-2$ spin-weighted spherical harmonics, $(\iota, \varphi)$ are the inclination and coalescence phase, $ \{ \alpha_{P \rightarrow I}, \beta_{ P \rightarrow I},  \gamma_{ P \rightarrow I} \} $ are Euler angles of the composite $P \to I$ rotation, which absorb the source angles $ \varphi $ and $ \iota $~\cite{RamosBuadesv5}, and $ h^{ P}_{\ell m} $ are the \emph{co-precessing modes}.

The modes in the \emph{P} frame take the form
\begin{equation}
\label{eq:h_bound_unbound}
\begin{aligned}
&h_{\ell m}^{ P} =
\left\{
\begin{array}{ll}
h_{\ell m}^{\text {F}},
& \text{unbound orbits},
\\
h_{\ell m}^{\text {insp-plunge }} = 
N_{\ell m} \, h_{\ell m}^{\text {F}},
& \text{bound orbits, } t \leq t_{\text {match }},
\\
h_{\ell m}^{\text {merger-RD }}\hspace{-25.5pt},
& \text{bound orbits, } t \geq t_{\text {match }},
\end{array}
\right.
\end{aligned}
\end{equation}
for $ (\ell, |m|) = \{(2,2), (3,3), (2,1), (4,4), (3,2), (4,3)\} $, where $  h_{\ell m}^{\text {F}} $ are the factorized EOB modes, $ N_{\ell m} $ are nonquasicircular corrections improving the late-inspiral waveform, and $ h_{\ell m}^{\text {merger-RD }} $ is the merger-ringdown portion starting at $ t _{ \text{match}} $ (see \EndMatter[em:imr]).
Unbound (bound) orbits are those with $H_{\text{EOB}} > \nolinebreak M$ ($H_{\text{EOB}} < M$); in dynamical captures, GW emission drives $H_{\text{EOB}}$ below $M$ and the bound prescription takes over.
The negative-$m$ modes are obtained via $ h _{ \ell \, -m} ^{ P} = (-1)^\ell h _{ \ell m} ^{ P *} $, thus neglecting mode asymmetries~\cite{Boyle:2014ioa}.

The factorized modes $ h _{ \ell m}^{\text {F}} $ resum analytical PN results~\cite{Damour:2007xr,Damour:2007yf,Damour:2008gu,Pan:2010hz} and are computed in the \emph{P}-frame using the \texttt{SEOBNRv6EHM} factorization $ h_{\ell m}^\mathrm{F}  = h_{\ell m}^\mathrm{F,\,hyb} \, h_{\ell m}^\mathrm{ecc} $, where $ h_{\ell m}^\mathrm{F,\,hyb} $ includes the leading-order terms for generic orbits and high-PN-order QC contributions, and $ h_{\ell m}^\mathrm{ecc} $ are eccentricity corrections at relative 1PN order.
These elements are given in Eqs.~(62) and (A22) of Ref.~\cite{Gamboa:2026jht}, and depend on the same variables $ (\Omega, r, \dot r, \dot p_{r_*, \s\text{cons}}) $ as the RR force.
The modes are evaluated on the solution of Eqs.~\eqref{eq:EOM}, and the spin contributions entering $ h _{ \ell m}^{\text {F}} $ are updated with the spin projections $ \bm{l}_{ \text{N}} \cdot \bm{\chi}_{i} $.

The merger-ringdown modes $ h_{\ell m}^{\text {merger-RD }} $ are built from a phenomenological QC prescription informed by Kerr quasinormal mode frequencies, taking into account the precession of the \emph{P} frame around $ \bm{J}_{ \text{f}} $~\cite{OShaughnessy:2012iol,Hamilton:2023znn,RamosBuadesv5} (see \EndMatter[em:imr]).

%%%%%%%%%%%%%%%%%%%%%%%%%%%
%%%%%%%%%%%%%%%%%%%%%%%%%%%
%%%%%%%%%%%%%%%%%%%%%%%%%%%

\paragraph{Agreement with NR waveforms}

\begin{figure}
\hspace{-5pt}
\includegraphics[width=\linewidth]{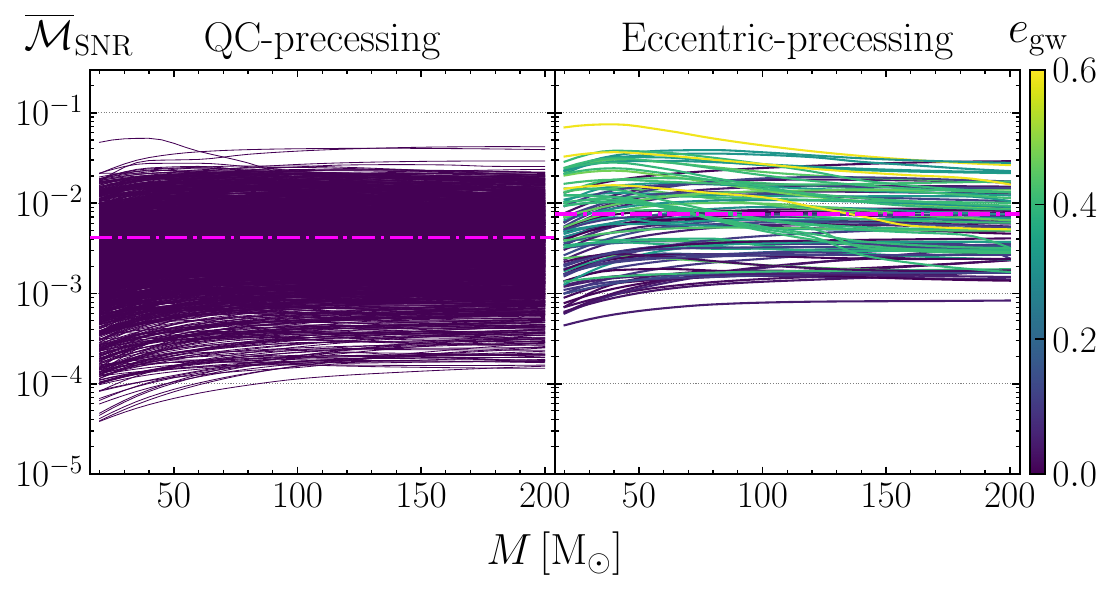}
\vspace{-5pt}
\caption{
SNR-weighted mismatches of \texttt{SEOBNRv6EPHM} against 1437 QC and 87 eccentric \texttt{SXS} NR waveforms of spin-precessing BBHs, at inclination $ \iota _{ \text{s}} = \pi/3 $.
Curve color indicates the initial GW eccentricity $ e_{\text{gw}} $ of each NR waveform; magenta dot-dashed lines mark the median of the per-simulation maximum mismatches.
}
\label{fig:mms_qc_ecc_v6}
\end{figure}

We quantify the accuracy of \texttt{SEOBNRv6EPHM} by comparing against 1437 QC and 87 eccentric NR waveforms of spin-precessing BBHs from the \texttt{SXS} collaboration~\cite{Scheel:2025jct} (eccentric simulations have mass ratio $ q \leq 6 $, effective precession spin~\cite{Schmidt:2014iyl} $ \chi _{ \text{p}} \leq 0.9 $, and initial GW eccentricities~\cite{Ramos-Buades:2022lgf,Shaikh:2023ypz,Shaikh:2025tae} $ e_{\text{gw}} \leq 0.6 $).
As a metric, we use a signal-to-noise-ratio (SNR)-weighted mismatch $ \overline{\mathcal{M}}_{\text{SNR}} $~\cite{Ossokine:2020kjp,Cotesta:2018fcv} (see \EndMatter[em:nr]) computed across the total mass range $ [20, \s 200] \, \solarmass $.
The main results are presented in Fig.~\ref{fig:mms_qc_ecc_v6}, where we plot the mismatches for all QC and eccentric NR waveforms, color-coded by their initial $ e_{\text{gw}} $ value.
We provide complementary information in the \SuppMat[sm:nr], including comparisons with \texttt{SEOBNRv5PHM} and \texttt{TEOBResumS-Dal\'i} (Figs.~\ref{fig:waveform_comparisons}--\ref{fig:hist_mms}) and the mismatch dependence on the source parameters (Fig.~\ref{fig:mismatches_corner}).
 
The median mismatch is below $ 1\% $ for both QC and eccentric cases, with challenging binaries reaching $ \lesssim 10\% $.
Systems with high eccentricities, strong precession, or large mass ratios are less accurately modeled, though the accuracy depends nontrivially on their combination.
Remarkably, binaries with $ q \leq 4 $ (as is the case for most observed BBHs~\cite{LIGOScientific:2026ctl}) consistently show mismatches below $ 4 \%$, even at high eccentricities, $ e_{\text{gw}} \sim 0.5 $;
the model inherits here the good performance in the aligned-spin limit~\cite{Gamboa:2026jht}.
Finally, \texttt{SEOBNRv6EPHM} matches the accuracy of \texttt{SEOBNRv5PHM} in the QC limit, and is more accurate than \texttt{TEOBResumS-Dal\'i} by a median factor of $ \sim 4 $ for both QC and eccentric systems, reaching more than an order of magnitude for several configurations (as is generically the case in the eccentric, aligned-spin limit~\cite{Gamboa:2026jht}).

The comparable accuracy for moderately eccentric and QC systems indicates that the spin-precessing baseline, not the eccentric sector, dominates the error budget:
as with the \texttt{SEOBNRv5} models~\cite{Pompiliv5,RamosBuadesv5,Estelles:2025zah}, the QC, spin-precessing accuracy does not reach the aligned-spin one~\cite{Gamboa:2026jht}.
Since precession mixes co-precessing modes of the same $ \ell $ but different $ m $~\cite{Schmidt:2010it}, mismodeling (or omission) of higher-order modes propagates into the full waveform.
Improving these modes~\cite{FooInPrep}, together with NR calibration of the spin-precessing sector, will be the focus of future work.

%%%%%%%%%%%%%%%%%%%%%%%%%%%
%%%%%%%%%%%%%%%%%%%%%%%%%%%
%%%%%%%%%%%%%%%%%%%%%%%%%%%

\paragraph{Agreement with NR scattering angles}

We validate the \texttt{SEOBNRv6EPHM} dynamics by comparing with recent, pioneering NR results for the scattering of generic-spin BHs from Ref.~\cite{Clark:2026bgg}:
equal-mass BHs with initial spins $ \bm{S}_1 = 0.125 \s \bm{\hat x} $ and $ \bm{S}_2 = \bm{0} $ and varying impact parameter $ b $.
The NR setup, the matching to EOB initial conditions, and the extraction of the azimuthal $ \Phi_\text{s} $ and polar $ \Theta_\text{s} $ scattering angles and Euler angles from the EOB trajectory are detailed in the \EndMatter[em:scattering].

\begin{figure}
\hspace{-5pt}
\includegraphics[width=\linewidth]{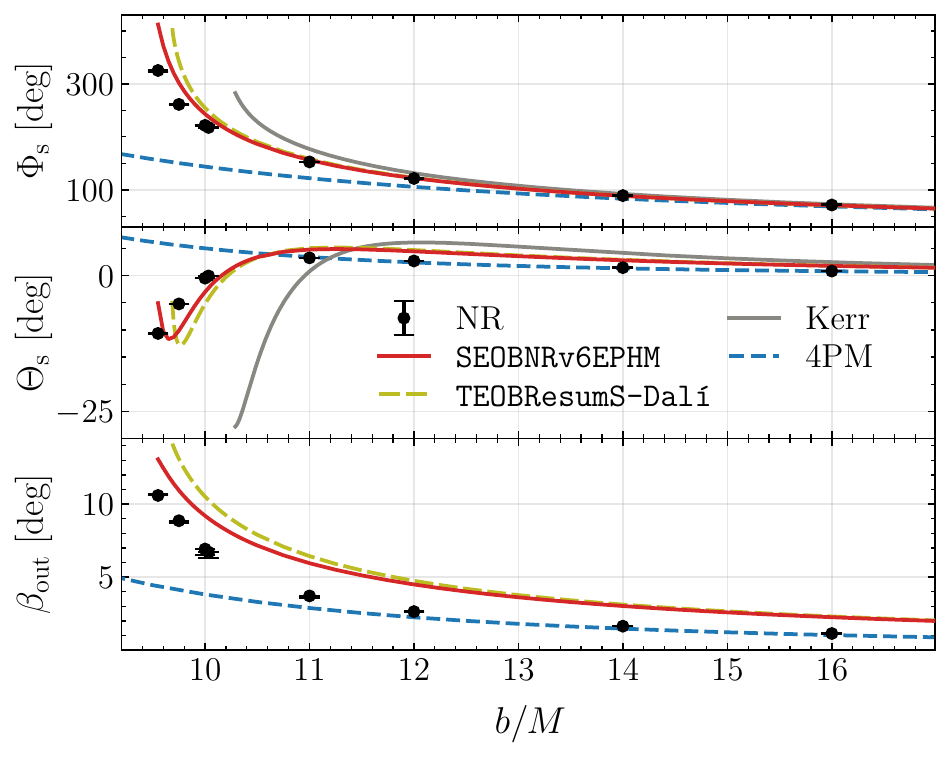}
\vspace{-5pt}
\caption{
Azimuthal $\Phi_\text{s}$ and polar $\Theta_\text{s}$ scattering angles and asymptotic Euler angle $\beta_\text{out}$ versus impact parameter $ b $, for equal-mass BHs with $\bm{S}_1 = 0.125 \s \bm{\hat x}$, $\bm{S}_2 \!=\! \bm{0}$, and $E_\text{ADM}/M=1.02281$, comparing both EOB models with NR, Kerr geodesics (top two panels), and 4PM results~\cite{Clark:2026bgg}.
\texttt{SEOBNRv6EPHM} is plotted down to the smallest NR $ b $; the \texttt{TEOBResumS-Dal\'i} curve ends where the model predicts capture.
}
\label{fig:scattering_angles}
\end{figure}

Figure~\ref{fig:scattering_angles} shows the scattering angles $\Phi_\text{s}$ and $\Theta_\text{s}$, and the asymptotic Euler angle $\beta_\text{out}$, predicted by NR, Kerr geodesics, fourth-order post-Minkowskian (PM) theory (all data from Ref.~\cite{Clark:2026bgg}), \texttt{SEOBNRv6EPHM}, and \texttt{TEOBResumS-Dal\'i}.
For $ \Phi _\text{s} $, EOB models fit NR best, with \texttt{SEOBNRv6EPHM} the closest: the azimuthal deflection is reasonably well modeled in the co-precessing frame.
In the weak field (large $ b $), PM estimates perform well, whereas both EOB models overpredict the polar deflection and final tilt of $ \bm l _{ \text{N}} $ (encoded in $\beta_\text{out}$), likely because the QC precession equations [Eqs.~\eqref{eq:sdot}--\eqref{eq:L}] omit radial contributions.
Toward the strong field, however, PM estimates depart significantly from NR for all three angles, and miss the turning point in $ \Theta_\text{s} $---a non-perturbative feature absent in PM expansions~\cite{Clark:2026bgg}.
Kerr geodesics and both EOB models reproduce it (as expected, since $ H_{\text{EOB}} $ includes exactly the Kerr Hamiltonian), with \texttt{SEOBNRv6EPHM} the closest to NR.

Near the smallest simulated $ b $, both EOB models predict a rise in $ \Theta_{\text{s}} $ as the separatrix between scattering and capture is approached.
This follows from the relation $ \tan \Theta_{\text{s}} =  \tan \beta_{\text{out}} \s \cos(\Phi_{\text{s}} - \alpha_{\text{out}}) $~\cite{Clark:2026bgg} between the scattering angles and the asymptotic Euler angles $ \alpha_{\text{out}} $ and $ \beta_{\text{out}} $:
as $ \Phi_{\text{s}} $ grows steeply near the separatrix, $ \Theta_{\text{s}} $ oscillates.
Testing this prediction will require NR simulations at smaller $ b $.

%

%%%%%%%%%%%%%%%%%%%%%%%%%%%
%%%%%%%%%%%%%%%%%%%%%%%%%%%
%%%%%%%%%%%%%%%%%%%%%%%%%%%

\paragraph{Analysis of GW200129}

We employ \texttt{SEOBNRv6EPHM} to perform the first joint eccentric, spin-precessing analysis of GW200129\_065458~\cite{LIGOScientific:2021djp} (hereafter GW200129).
The QC, spin-precessing model \texttt{NRSur7dq4}~\cite{Varma:2019csw} yielded evidence for spin precession~\cite{Hannam:2021pit}, while eccentric, aligned-spin models favored nonzero eccentricity over the QC hypothesis~\cite{Gupte:2024jfe, Planas:2025jny, Malagon:2026uev, Pompili:2026yxq}.
Both interpretations are complicated by a glitch overlapping the signal in LIGO-Livingston: the evidence for precession~\cite{Payne:2022spz,Hoy:2026dkr,Cheung:2026myt} and eccentricity~\cite{Gupte:2024jfe} depends on its treatment (see the \SuppMat[sm:gw200129] for a summary).
Motivated by this, we run eccentric, spin-precessing analyses under six glitch mitigations---none, \texttt{gwsubtract}~\cite{LIGOScientific:2021djp}, nonlinear noise subtraction (NLSUB)~\cite{Macas:2023wiw}, and three fair \texttt{BayesWave} draws (A--C) from Ref.~\cite{Payne:2022spz}---each with a uniform ($e\in[0,0.8]$) and a log-uniform ($e\in[10^{-4},0.8]$) eccentricity prior.

\begin{figure}
\hspace{-5pt}
\includegraphics[width=\linewidth]{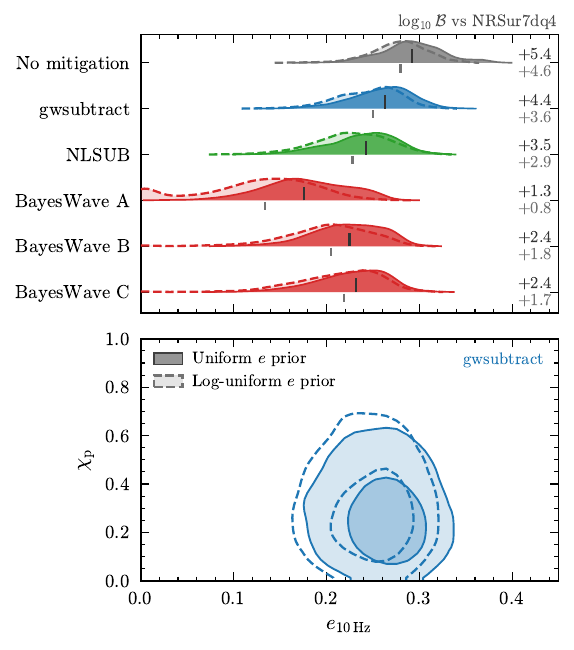}
\vspace{-5pt}
\caption{
Parameter estimation of GW200129 with \texttt{SEOBNRv6EPHM}.
\emph{Top}: posterior on the eccentricity at 10~Hz for the six treatments of the LIGO-Livingston glitch described in the text; bars mark the medians.
\emph{Bottom}: joint posterior on $e_{10\,\text{Hz}}$ and $\chi_{\text{p}}$ for the \texttt{gwsubtract} data, with contours enclosing $50\%$ and $90\%$ of the posterior.
In both panels, dark filled distributions with solid outlines assume a uniform prior on $e$, and lighter dashed ones a log-uniform prior.
Numbers on the right give $\log_{10}\mathcal{B}$ against the QC precessing model \texttt{NRSur7dq4} for each prior (uniform above, log-uniform below).
}
\label{fig:gw200129_eccentricity}
\end{figure}

Eccentricity is favored over the QC precessing hypothesis (\texttt{NRSur7dq4}) under all six treatments (see Fig.~\ref{fig:gw200129_eccentricity}, and Table~\ref{tab:gw200129_table} of the \SuppMat[sm:gw200129]), with $\log_{10}\mathcal{B}\simeq0.8$--$5.4$ and $e_{10\,\text{Hz}}\simeq0.13$--$0.29$; the support is largest for the unmitigated, \texttt{gwsubtract}, and NLSUB data and smallest for the \texttt{BayesWave} draws.
With eccentricity included, no treatment strongly supports precession ($\log_{10}\mathcal{B}\lesssim 0.5$), and $\chi_{\text{p}}$ is weakly constrained.
The model omits mode asymmetries in the co-precessing frame, which can affect precession measurements~\cite{Kolitsidou:2024vub, Estelles:2025zah}.
Nevertheless, injection-recovery studies of two synthetic eccentric, spin-precessing NR signals---one with properties close to those inferred for GW200129, the other more eccentric and more strongly precessing---show that \texttt{SEOBNRv6EPHM} recovers their source parameters without significant bias, even at network SNRs above that of GW200129 (see the \SuppMat[sm:injections]).
This NR-injection validation---to our knowledge, the first for an eccentric, spin-precessing model---indicates that the recovered eccentricity is robust to waveform systematics, strengthening the eccentric interpretation of GW200129.

Ten additional GW events, with a focus on previously reported eccentric candidates, are analyzed in the \SuppMat[sm:events] (Table~\ref{tab:real_events_table} and Fig.~\ref{fig:ecc_chip_corner_grid}), yielding results broadly consistent with eccentric, aligned-spin analyses~\cite{Pompili:2026yxq}.

%%%%%%%%%%%%%%%%%%%%%%%%%%%
%%%%%%%%%%%%%%%%%%%%%%%%%%%
%%%%%%%%%%%%%%%%%%%%%%%%%%%

\paragraph{Computational efficiency}

We quantify the speed of \texttt{SEOBNRv6EPHM} via PE analyses of representative GW events.
Table~\ref{tab:benchmarks} shows the runtime and number of likelihood evaluations for
GW150914 (an event with typical BBH parameters), GW190521 (a short-duration signal), GW200105\_162426 (hereafter GW200105, a long neutron-star--BH binary event), and GW170817 (a long binary neutron star signal)
analyzed with standard settings (see \EndMatter[em:pe_settings]) using \texttt{SEOBNRv5PHM} and \texttt{SEOBNRv6EPHM} (with and without ($ e = 0 $) sampling over $ e $ and $ 
\zeta $).
Comparisons with \texttt{TEOBResumS-Dal\'i} and additional benchmarks are presented in the \SuppMat[sm:benchmarks].

Eccentric, spin-precessing analyses with \texttt{SEOBNRv6EPHM} are faster than QC analyses with \texttt{SEOBNRv5PHM} for all but the longest signals, which are at most $\sim1.5$ times slower;
in the $ e=0 $ case, \texttt{SEOBNRv6EPHM} is $ 1.7 $--$ 3.3 $ times faster, while recovering posterior distributions consistent with \texttt{SEOBNRv5PHM}.
Against \texttt{TEOBResumS-Dal\'i}, \texttt{SEOBNRv6EPHM} is always faster, with speed-ups that grow with signal duration, reaching a factor of $ \sim 10 $ for the longest events.
Since \texttt{SEOBNRv5PHM} is widely employed, e.g., by LVK studies~\cite{LIGOScientific:2025slb,LIGOScientific:2025yae,LIGOScientific:2026ifv,LIGOScientific:2026wfs}, our results show that large-scale GW analyses can now incorporate joint eccentric, spin-precessing inference at comparable computational cost.

\begin{table}
\caption{
Runtime and likelihood $\mathcal{L}$ evaluations (in units of $10^7$) of the PE analyses of representative GW events, using \texttt{SEOBNRv5PHM} and \texttt{SEOBNRv6EPHM} in its QC and eccentric configurations.
}
\input{tab/benchmarks}
\vspace{-5pt}
\label{tab:benchmarks}
\end{table}
%

%%%%%%%%%%%%%%%%%%%%%%%%%%%
%%%%%%%%%%%%%%%%%%%%%%%%%%%
%%%%%%%%%%%%%%%%%%%%%%%%%%%

\paragraph{Conclusions}

We presented \texttt{SEOBNRv6EPHM}:
an accurate and efficient model for generic-orbit BBHs.
It is built upon the aligned-spin model \texttt{SEOBNRv6EHM}~\cite{Gamboa:2026jht}, and hence calibrated only to QC, aligned-spin NR simulations.
Median mismatches against spin-precessing NR waveforms are below $ 1\% $, with maxima below $ 4\% $ for binaries with $ q \leq 4 $ even at $ e_{\text{gw}} \sim 0.5 $ and $ \chi_{\text{p}} \sim 0.8 $.
The model matches the accuracy of \texttt{SEOBNRv5PHM} and improves on \texttt{TEOBResumS-Dal\'i} by a median factor of $ 4 $.
\texttt{SEOBNRv6EPHM} reproduces the non-perturbative behavior observed in NR simulations of generic-spin BH scattering~\cite{Clark:2026bgg} more accurately than \texttt{TEOBResumS-Dal\'i}.
The BBH generic-orbit sector is now described within a single \texttt{SEOBNR} framework, opening the way to improved studies of generic-orbit GW signals~\cite{Pompili:2026yxq,Lange:2026eqx}.

\texttt{SEOBNRv6EPHM} enables, for the first time, accurate and computationally feasible large-scale GW analyses of generic BBH configurations.
We demonstrated this with the first eccentric, spin-precessing analyses of GW200129 under different treatments of the glitch overlapping the signal: all favor eccentricity over the QC hypothesis.
These results are supported by zero-noise injections of synthetic eccentric, spin-precessing NR signals, whose source parameters are recovered without significant bias.
We ran eccentric, spin-precessing PE analyses on ten additional GW events and dedicated benchmarks on representative cases (including long binary neutron star signals).
Importantly, we found that eccentric, spin-precessing PE with \texttt{SEOBNRv6EPHM} is faster than QC PE with \texttt{SEOBNRv5PHM} for typical GW events, and it can be up to an order of magnitude faster than \texttt{TEOBResumS-Dal\'i} for the longest durations. 
Thus, it is now realistic to analyze GW catalogs accounting for eccentricity and spin precession (the key signatures of dynamical formation), and to perform systematic studies of PE biases in preparation for the high-SNR observations of future GW detectors such as LISA~\cite{LISA:2024hlh}, the Einstein Telescope~\cite{ET:2025xjr}, and Cosmic Explorer~\cite{Evans:2021gyd}.

Future work includes modeling mode asymmetries (enabling studies of GW kicks in generic BBHs), improving the co-precessing higher-order modes, adding eccentric terms to the precession equations, incorporating insights from the test-mass limit~\cite{Albanesi:2021rby,Albanesi:2022ywx,Albanesi:2023bgi,Faggioli:2024ugn,Islam:2024vro,Islam:2025wci,Faggioli:2025hff,Faggioli:2026alx,Nishimura:2026nse}, calibrating to eccentric, spin-precessing NR simulations, and accelerating the model with near-identity averaging transformations~\cite{Lynch:2026ibo}.

%%%%%%%%%%%%%%%%%%%%%%%%%%%
%%%%%%%%%%%%%%%%%%%%%%%%%%%
%%%%%%%%%%%%%%%%%%%%%%%%%%%

\paragraph{Acknowledgments}

We are grateful to Adam Clark, Geraint Pratten, and Patricia Schmidt for sharing their BH scattering data.
We thank Teagan A.~Clarke for reviewing this manuscript within the LIGO Scientific Collaboration.
We benefited from discussions with
Aurora Abbondanza,
H\'ector Estell\'es,
Guglielmo Faggioli,
Cheng Foo,
Nihar Gupte,
Marcus Haberland,
Mohammed Khalil,
Oliver Long,
Philip Lynch,
Maarten van de Meent,
Gonzalo Morras,
Peter James Nee,
Nami Nishimura,
and
Lluc Planas.

The authors acknowledge the use of \texttt{Claude}~\cite{claude_opus_4_8} for code optimizations.
The computational work for this manuscript was carried out on the \texttt{Hypatia} computer cluster at the Max Planck Institute for Gravitational Physics in Potsdam.
Numerical-relativity simulations were performed at the Max Planck Computing and Data Facility on the Minerva HPC system of the ``Astrophysical and Cosmological Relativity'' division, and at the Dutch national e-infrastructure with the support of the SURF Cooperative (NWO-2024.002), and in Picasso Supercomputer from the Red Espa\~nola de Supercomputaci\'on (RES) (AECT-2025-1-0017, AECT-2025-2-0004, AECT-2025-3-0015, AECT-2026-1-0004).
This work is funded by the European Union (ERC grant GWSky/101167314).
Views and opinions expressed are however those of the authors only and do not necessarily reflect those of the European Union or the European Research Council Executive Agency. Neither the European Union nor the granting authority can be held responsible for them.
L.~P. is supported by a UKRI Future Leaders Fellowship (grant number MR/Y018060/1).
A.~R.-B. is supported by the Veni research programme which is (partly) financed by the Dutch Research Council (NWO) under the grant VI.Veni.222.396; acknowledges support from the Spanish Agencia Estatal de Investigaci\'on grant PID2024-157460NA-I00; and the Spanish Ministerio de Ciencia, Innovaci\'on y Universidades (Beatriz Galindo, BG23/00056), co-financed by UIB.
This material is based upon work supported by NSF's LIGO Laboratory which is a major facility fully funded by the National Science Foundation.
This research has made use of data or software obtained from the Gravitational Wave Open Science Center (gwosc.org), a service of LIGO Laboratory, the LIGO Scientific Collaboration, the Virgo Collaboration, and KAGRA. LIGO Laboratory and Advanced LIGO are funded by the United States National Science Foundation (NSF) as well as the Science and Technology Facilities Council (STFC) of the United Kingdom, the Max-Planck-Society (MPS), and the State of Niedersachsen/Germany for support of the construction of Advanced LIGO and construction and operation of the GEO600 detector. Additional support for Advanced LIGO was provided by the Australian Research Council. Virgo is funded, through the European Gravitational Observatory (EGO), by the French Centre National de Recherche Scientifique (CNRS), the Italian Istituto Nazionale di Fisica Nucleare (INFN) and the Dutch Nikhef, with contributions by institutions from Belgium, Germany, Greece, Hungary, Ireland, Japan, Monaco, Poland, Portugal, Spain. KAGRA is supported by Ministry of Education, Culture, Sports, Science and Technology (MEXT), Japan Society for the Promotion of Science (JSPS) in Japan; National Research Foundation (NRF) and Ministry of Science and ICT (MSIT) in Korea; Academia Sinica (AS) and National Science and Technology Council (NSTC) in Taiwan.

%%%%%%%%%%%%%%%%%%%%%%%%%%%
%%%%%%%%%%%%%%%%%%%%%%%%%%%
%%%%%%%%%%%%%%%%%%%%%%%%%%%

\paragraph{Data availability}

\texttt{SEOBNRv6EPHM} is implemented in the \texttt{pyseobnr} package~\cite{Mihaylovv5} and will be made publicly available after its upcoming review within the LVK collaboration.
We provide an ancillary file that lists the NR simulations used in this work, the binary parameters, and the mismatch information for each model.
Posterior samples for the GW events analyzed with \texttt{SEOBNRv6EPHM} are available at \url{https://doi.org/10.5281/zenodo.22065790}.

%%%%%%%%%%%%%%%%%%%%%%%%%%%
%%%%%%%%%%%%%%%%%%%%%%%%%%%
%%%%%%%%%%%%%%%%%%%%%%%%%%%

%\appendix

%\section{Appendix}

\onecolumngrid

\section{End Matter}
\label{sec:end_matter}

\twocolumngrid

%%%%%%%%%%%%%%%%%%%%%%%%%%%
%%%%%%%%%%%%%%%%%%%%%%%%%%%

\paragraph{Details of IMR-waveform construction}
\phantomsection\label{em:imr}

Building IMR waveforms requires identifying reference times along both the main dynamics and an aligned-spin background QC dynamics, applying phenomenological nonquasicircular (NQC) corrections to the analytical EOB modes, and using a phenomenological ansatz for the merger-ringdown.

The background QC dynamics uses the spins $ \bm{l} _{ \text{N}} \cdot \bm{\chi}_{i} $ from the eccentric dynamics at the time $ t _{ \text{ref,\s\s s}} $ defined by $ H _{ \text{EOB}} ^{ \text{pprec}}(t _{ \text{ref,\s\s s}}) =  E _{ \text{c}} ^{ 2\text{PN, noS}}\s \big(\s \Omega _{ \text{c}}^{ 2\text{PN, noS}} |_{r = 9M} \big) $, where $ E _{ \text{c}} ^{ 2\text{PN, noS}} $ and $ \Omega _{ \text{c}}^{ 2\text{PN, noS}} $ are nonspinning 2PN expressions for the energy and frequency of circular orbits (Eq.~(B8a) of Ref.~\cite{Gamboa:2024imd} and Eq.~(A24) of Ref.~\cite{Gamboa:2026jht}), evaluated at $ r = 9M $ as this point always exists in a QC inspiral.
Since energy decreases monotonically, $ t _{ \text{ref,\s\s s}} $ is unique and robust under parameter perturbations; if the eccentric dynamics starts below this energy, we integrate backward.

The matching time $ t _{ \text{match}} $ between the inspiral-plunge and merger-ringdown portions is calculated using the QC dynamics and a QC-NR-calibrated parameter.
First, the eccentric and QC dynamics are aligned at the times $ t _{ \text{ref,\,ecc}} $ and $ t _{ \text{ref,\,qc}} $ at which each reaches the larger of the two final separations, $ \max ( r_\text{final,\,ecc}, \, r _{ \text{final,\,qc}}) $.
Then, we compute
$ t _{ \text{match}} \equiv \nolinebreak t _\text{ISCO} ^\text{ecc} + \Delta t_\text{ISCO} ^{ 22} $,
where
$ t _\text{ISCO} ^\text{ecc} \equiv t _\text{ISCO} ^\text{qc} + t _{ \text{ref,\,ecc}} - t _{ \text{ref,\,qc}} $.
Here, $ t _\text{ISCO} ^{\text{qc}^{\phantom{x}}} $ is the time at which the QC separation equals the Kerr innermost stable circular orbit (ISCO) radius~\cite{Bardeen:1972fi}, calculated with the spin projections $ \bm{l} _{ \text{N}} \cdot \bm{\chi}_{i} $ at $ t _{ \text{ref,\s\s s}} $, and $ \Delta t_\text{ISCO} ^{ 22} $ is a parameter calibrated to aligned-spin QC NR simulations, given in Eq.~(B3) of Ref.~\cite{Gamboa:2026jht} and evaluated in terms of $ (\nu, \bm{l}_{ \text{N}} \cdot \bm{a}_{\pm} ) $ at $ t_\text{ISCO} ^\text{ecc} $.
This construction is consistent with \texttt{SEOBNRv6EHM} and transfers the QC NR calibration to generic systems.

To improve the late-inspiral waveform, we apply NQC corrections $ N _{ \ell m} $~\cite{Damour:2007vq,Damour:2008te,Buonanno:2009qa,Bohe:2016gbl}, which enforce agreement of amplitudes and frequencies of the modes with fits to aligned-spin QC NR simulations~\cite{Pompiliv5}.
We evaluate the NQC corrections (given in Eq.~(73) of Ref.~\cite{Gamboa:2026jht}) on the QC dynamics, confining their effect to the end of the inspiral while retaining the computational efficiency of an aligned-spin system.

Following \texttt{SEOBNRv5PHM}~\cite{RamosBuadesv5}, the merger-ringdown modes $ h_{\ell m}^{\text {merger-RD }} $ are obtained from a phenomenological QC ansatz (Eq.~(78) of Ref.~\cite{Gamboa:2026jht}) representing damped oscillations of the remnant Kerr BH.
The quasinormal mode (QNM) frequencies~\cite{Berti:2005ys} are first calculated in the \emph{J} frame (defined by $ \bm{J}_{ \text{f}} \equiv (\bm{L} + \bm{S}_{1} + \bm{S}_{2}) |_{t = t_\text{match}} $), using QC NR-informed fits~\cite{Jimenez-Forteza:2016oae, Hofmann:2016yih} for the final mass $ M _{ \text{f}} $ and spin magnitude $ \chi_{ \text{f}} $ of the remnant, evaluated at $ t _{ \text{ref,\s\s s}} $.
The \emph{P}-frame QNM frequencies are then obtained with a shift that accounts for the precession of the \emph{P} frame around $ \bm{J}_{ \text{f}} $~\cite{OShaughnessy:2012iol,Hamilton:2023znn,RamosBuadesv5}.

%%%%%%%%%%%%%%%%%%%%%%%%%%%
%%%%%%%%%%%%%%%%%%%%%%%%%%%

\paragraph{Details of NR waveform comparisons}
\phantomsection\label{em:nr}

The \emph{strain} in a GW detector can be written as
\begin{equation}
\label{eq:h}
h(t) =
\mathcal{A} \left ( h_+ \cos \kappa + h_\times \sin \kappa\right ) ,
\end{equation}
where $ \mathcal{A} $ and $ \kappa $ are an effective amplitude and polarization angle which depend on the source's sky location in the detector frame~\cite{Cotesta:2018fcv,Ossokine:2020kjp}, and $ h_{+, \times} $ are the polarizations which depend on the inclination $ \iota $ and azimuthal angle $ \varphi $ of the line-of-sight in the source frame, the luminosity distance $d_L$, the coalescence time $t _{ \text{c}}$, and a set of intrinsic parameters $ \bm{\Pi} $ that determine a unique physical binary.
For generic binaries, $ \bm{\Pi} = \{ m_1, m_2, \bm{S}_1, \bm{S}_2 \} \cup \bm{\Pi} ^{ \text{ecc}} $, where $ \bm{\Pi} ^{ \text{ecc}} = \{ e, \zeta \} $ for bound orbits, with $ e $ and $ \zeta $ (a generic radial phase parameter) specified, e.g., at an orbit-averaged frequency $ \langle M \Omega \rangle $, or $ \bm{\Pi} ^{ \text{ecc}} = \{ E, J _{ \perp} \} $ specified at a separation $ r $ for generic orbits.

A basic element in waveform comparisons is the \emph{overlap} between two strains $ h_1 $ and $ h_2 $, defined as the inner product
$ \langle h_1 | \, h_2 \rangle \equiv 4 \,\Re \int^{f_{\text{max}}}_{f_{\text{min}}} \tilde{h}_1 \,\tilde{h}_2 {\!}^* / S_\text{\!n} \, \text{d}f $~\cite{Sathyaprakash:1991mt,Finn:1992xs},
where tildes denote Fourier transforms, stars complex conjugation, and $S_\text{\!n}$ is the one-sided power-spectral density (PSD) of the detector's noise.
In this work, we taper the time-domain strains with a Planck window~\cite{McKechan:2010kp}, use the \texttt{A+} PSD of LIGO detectors~\cite{sensitivity_curves}, and set $f_{\text{min}}=10\,$Hz and $f_{\text{max}}=2048\,$Hz.
For signals starting above $10\,$Hz, we instead set $f_{\text{min}} = 1.35 \, f_{\text{start}}$, with $f_{\text{start}}$ the initial orbit-averaged frequency of the co-precessing $(2,2)$ mode, to mitigate Fourier-transform artifacts.

The agreement between signal $ h _{ \text{s}} $ (NR) and template $ h _{ \text{t}} $ (EOB) is quantified with the sky-and-polarization-averaged, SNR-weighted mismatch~\cite{Ossokine:2020kjp,Cotesta:2018fcv}
\begin{equation}
\label{eq:mismatch}
\overline{\mathcal{M}}_{\text{SNR}}(M, \iota _{ \text{s}}) \equiv 1 - \sqrt[3]{\frac{\int_{0}^{2\pi} d\varphi_{\text{s}} \int_{0}^{2\pi} d\kappa_ {\text{s}} \ \mathcal{F}^{3} \ \text{SNR}^3}{\int_{0}^{2\pi} d\varphi_{\text{s}} \int_{0}^{2\pi} d\kappa_{\text{s}} \ \text{SNR}^3}},
\end{equation}
where $ \mathcal{F} $ is the template's \emph{faithfulness} and $ \text{SNR} \equiv \sqrt{\langle h_{\text{s}} | \s h_{\text{s}} \rangle} $.
The faithfulness $ \mathcal{F} $ is a \emph{maximized overlap},
$ \mF (M, \Sigma _{ \text{s}}) = \max_{\Sigma _{ \text{t}}} \, \big \langle \hat h _{ \text{s}} | \s \hat h _{ \text{t}} \big \rangle $,
of normalized strains
\big($\hat h_{ \text{s,\s t}} \equiv h_{ \text{s,\s t}} / \! \sqrt{ \langle h_{ \text{s,\s t}} \s | \s h_{ \text{s,\s t}} \rangle }$\s\big),
in which some template parameters $ \Sigma _{ \text{t}} $ are optimized for a total mass $ M $ and source parameters $ \Sigma _{ \text{s}} $.
The maximizations are justified since the optimized parameters either have limited astrophysical relevance or are defined differently across models.

For QC-waveform comparisons, we optimize analytically over the effective polarization angle $ \kappa _{ \text{t}} $~\cite{Capano:2013raa,Harry:2016ijz,Harry:2017weg} and numerically over the coalescence time $ t _{ \text{c,\s t}} $, azimuthal phase $ \varphi _{ \text{t}} $, and a solid rotation $ \sigma $ of the in-plane spin angle $ \phi_{12,\s \text{s}} = \phi_{12,\s \text{t}}$~\cite{RamosBuadesv5,Estelles:2025zah}.
For eccentric waveforms, we additionally optimize over the eccentricity $ e _{ \text{t}} $, radial phase $ \zeta _{ \text{t}} $, and starting frequency $ \langle M \Omega \rangle _{ \text{t}} $ (these could instead be estimated from the NR waveform~\cite{Shaikh:2023ypz,Shaikh:2025tae}, but such estimation is not error-free) with a differential evolution algorithm~\cite{Storn:1997uea,Morras:2026fho} (we use the implementation in \texttt{SciPy}~\cite{2020SciPy-NMeth} with \texttt{popsize}=$35$, \texttt{maxiter}=$300$, \texttt{rtol}=$10^{-3}$, and \texttt{atol}=$10^{-5}$).
To ensure the comparison uses most of the signal, we require the times to merger of signal and template to differ by no more than $ 0.7\% $.
To set the parameter bounds, we first estimate the signal's initial GW eccentricity $ e _{ \text{gw},\s 0} $ and orbit-averaged GW frequency $ \omega_{ \text{gw},\s 0} $ with \texttt{gw\_eccentricity}~\cite{Shaikh:2023ypz,Shaikh:2025tae}.
The eccentricity and starting frequency bounds are then taken as
$ [e _{ \text{min}}, e _{ \text{max}}] = [e _{ \text{gw},\s 0} - 0.2, e _{ \text{gw},\s 0} + 0.2] $ (we enforce $ e _{ \text{min}} \geq \nolinebreak 0 $ and $ e _{ \text{max}} \leq 0.9 $),
and
$ [\langle M \Omega \rangle _{ \text{min}}, \langle M \Omega \rangle _{ \text{max}}] = [ \langle M \Omega \rangle _{e _\text{max}}, \langle M \Omega \rangle _{ e _\text{min}}] $, respectively.
Here, $ \langle M \Omega \rangle _{e _\text{max}} $ and $ \langle M \Omega \rangle _{e _\text{min}} $ are the frequencies at which the template (with $ e _{ \text{t}} = e _\text{max} $ or $ e _\text{min} $) matches the signal's time to merger, and are found by doing a starting-frequency root-search seeded with $ \omega_{ \text{gw},\s 0}/2 $ and using $ \varphi _{ \text{t}} = \sigma = \zeta _{ \text{t}} = 0$.
Because increasing the eccentricity shortens the time to merger, the optimal frequency necessarily lies within these bounds.
Identical optimization routines are applied to every model, so cross-comparisons are unaffected.

The integrals in Eq.~\eqref{eq:mismatch} are computed as a discrete sum over a uniform grid of $ 4 \kern-0.08em \times \kern-0.08em 4 $ values for the signal angles $ \varphi_{\textrm{s}}, \kappa_{\text{s}} \in \nolinebreak{ [0, \s 2 \pi]} $ with $ \iota_{\text{s}} = \pi/3 $, for total masses $ M $ in the range $ [20, 200] \, \solarmass $ (template parameters are optimized at each total mass);
these settings allow direct comparisons with Ref.~\cite{RamosBuadesv5}.
For NR waveforms, we use all the inertial modes (without memory contributions~\cite{Scheel:2025jct}) up to $\ell = 4$.
For \texttt{SEOBNRv5PHM} and \texttt{SEOBNRv6EPHM}, we use the co-precessing modes $( \ell, |m| ) \in \nolinebreak \{(2, 2),\, (3, 3),\, (2, 1),\, (4, 4),\, (3, 2),\, (4, 3)\}$, with mode asymmetries in the $(2, 2)$, $(3, 3)$, and $(4, 4)$ modes for \texttt{SEOBNRv5PHM}~\cite{Estelles:2025zah}.
For \texttt{TEOBResumS-Dal\'i}~\cite{Gamba:2024cvy,Albanesi:2025txj}, we use $( \ell, |m| ) \in \nolinebreak \{(2, 2),\, (3, 3),\, (2, 1),\, (4, 4)\}$ as these have been reviewed (in the aligned-spin limit~\cite{Nagar:2024dzj}) by the LVK collaboration, and they were used in the analyses of Refs.~\cite{Gamba:2025qfg,Chandra:2025jfc,Albanesi:2025txj}.
We use the code with tag \texttt{v1.1.2-Dali} from \url{https://bitbucket.org/teobresums/teobresums}.

We employ 1437 QC (publicly available) and 87 eccentric ($ 62 $ private + $ 25 $ publicly available) NR waveforms of spin-precessing BBHs produced with the Spectral Einstein code (\texttt{SpEC})~\cite{SpECwebsite} from the \texttt{SXS} collaboration~\cite{Scheel:2025jct}.
The QC waveforms correspond to those used in the validation of the \texttt{SEOBNRv5PHM} model with mode asymmetries~\cite{Estelles:2025zah} (we removed deprecated simulations as per the latest release of the \texttt{SXS} catalog).
This enables a direct comparison with previous \texttt{SEOBNR} validations~\cite{RamosBuadesv5,Estelles:2025zah};
a systematic study of waveform-model accuracy over the full, extended \texttt{SXS} catalog~\cite{Scheel:2025jct} will be presented in upcoming work.
The $ 25 $ public eccentric simulations satisfy $ e _{ \text{gw}} > 0.005 $ and $ |\chi_{1x}| + |\chi_{1y}| + |\chi_{2x}| + |\chi_{2y}| > 10^{-3} $ at the relaxation time.
We provide an ancillary file listing the simulation identifiers, mass ratio $ q $, dimensionless spin vectors $ \bm{\chi}_1 $ and $ \bm{\chi}_2 $, initial GW eccentricity $ e _{ \text{gw}} $, effective spin
$\chi_\text{eff} = (\bm{a}_1 + \bm{a}_2) \cdot \bm l _{ \text{N}}/ M $,
and effective precession spin
$\chi_\text{p} = \max \big( \chi_{1\perp}, \chi_{2\perp} (4 + 3 q)/(4 q^2 + 3 q) \big)$~\cite{Schmidt:2014iyl},
where $\chi_{i\perp}$ are the magnitudes of the dimensionless spins projections onto the orbital plane.
These quantities are measured in a frame whose $ z $-axis is aligned with the orbital angular momentum at the relaxation time.
The numerical error of the NR waveforms is estimated as the mismatch $ \overline{\mathcal{M}}_{\text{SNR}} $ between the highest- and second-highest-resolution simulations, optimizing analytically over $ \kappa _{ \text{t}} $ and numerically over $ t _{ \text{c,\s t}} $ and $ \varphi _{ \text{t}} $.

%%%%%%%%%%%%%%%%%%%%%%%%%%%
%%%%%%%%%%%%%%%%%%%%%%%%%%%

\paragraph{Scattering setup and angle extraction}
\phantomsection\label{em:scattering}

We use the recent NR data from Ref.~\cite{Clark:2026bgg} produced with the \texttt{Einstein} \texttt{Toolkit}~\cite{EinsteinToolkit:2025}:
equal-mass BHs placed at a separation $ D = 100 M $, with initial spins $ \bm{S}_1 = 0.125 \s \bm{\hat x} $ and $ \bm{S}_2 = \bm{0} $ (orthogonal to $ \bm l _{ \text{N}} $), Arnowitt--Deser--Misner (ADM) linear momenta $ | \bm{P} _{ \text{ADM}} | = 0.11456 $ (giving an ADM energy $ E _{ \text{ADM}}/M = 1.02281 $), and $ \bm L_{ \text{N}} = b \, | \bm{P} _{ \text{ADM}} | \s \bm{\hat z} $, with different impact parameters $ b $.
For the EOB initial conditions, we use $ r = D $, $ E = E _{ \text{ADM}} $, and $ J _\perp = \bm{\hat z} \cdot \bm{J} = L_{ \text{N}, \s z} = b \, | \bm{P} _{ \text{ADM}} | $, with $J_\perp/M^2 = \nu \, p_\phi/(\mu M) + m_1^2/M^2 \, \bm{l}_{ \text{N}} \cdot \bm{\chi}_1 + m_2^2/M^2 \, \bm{l}_{ \text{N}} \cdot \bm{\chi}_2 $.
The 3D EOB trajectory $ (x, y, z) $ is obtained by rotating the co-precessing dynamics $ (r \cos \phi, r \sin \phi) $ with a quaternion tracking the $ \bm l _{ \text{N}} $ evolution.
We then extract the azimuthal and polar angles as $ \Phi(t) = \arctan(y /x) $ and $ \Theta(t) = \arcsin(z/r) $, with $ \Phi = \Theta = 0 $ at $ t = 0 $.
The scattering angles are calculated as $ \Phi _{\text{s}} \equiv \Phi _{ \text{out}} - \Phi _{ \text{in}} - \pi $ and $ \Theta_{\text{s}} \equiv \Theta _{ \text{out}} - \Theta _{ \text{in}} $, where the in/outgoing branches are obtained by extrapolating the angles at $ r \to \infty $ using least-squares polynomial fits in $ 1/r $ (we use polynomial orders $2$--$6$ over $ r > 30 M $, and take the mean of the extrapolated values).
The asymptotic Euler angles are obtained as $ \alpha _{\text{out}} = \arctan ( l _{ \text{N}, y} ^{ \text{out}} / l _{ \text{N}, x}^{ \text{out}} ) $ and $ \beta _{\text{out}} = \arccos ( l _{ \text{N},\s z}^{ \text{out}} ) $, using the $ xyz $-components of the extrapolated $ \bm l _{ \text{N}}^{ \text{out}} $ along the outgoing branch.

%%%%%%%%%%%%%%%%%%%%%%%%%%%
%%%%%%%%%%%%%%%%%%%%%%%%%%%

\paragraph{Settings of benchmark PE analyses}
\phantomsection\label{em:pe_settings}

The three configurations compared in Table~\ref{tab:benchmarks}---\texttt{SEOBNRv5PHM}, \texttt{SEOBNRv6EPHM} in its QC limit ($ e = 0 $), and the full eccentric, spin-precessing \texttt{SEOBNRv6EPHM}---share identical data, priors, and sampler settings, so that the quoted wall-clock runtimes differ primarily through the cost of the waveform model.
We follow the data, prior, frequency, and sampler settings of Ref.~\cite{Pompili:2026yxq}, except for the spin prior, which is taken to be isotropic in orientation and uniform in magnitude.
Strain data, PSDs, and calibration envelopes are obtained from the Gravitational Wave Open Science Center (GWOSC)~\cite{LIGOScientific:2019lzm, KAGRA:2023pio}, and inference is done with \texttt{Bilby}~\cite{Ashton:2018jfp, Romero-Shaw:2020owr} and the \texttt{dynesty} sampler~\cite{Speagle:2019ivv}, using the \texttt{acceptance-walk} method with \texttt{naccept}=$60 $, \texttt{nlive}=$1000 $, and distance marginalization.
These choices are close to those adopted in the latest LVK analyses~\cite{LIGOScientific:2026ifv}.
The two BBH signals are analyzed on a single node with 64 CPU cores, while the longer GW200105 and GW170817 signals use \texttt{parallel Bilby}~\cite{Smith:2019ucc} across 16 nodes with 32 cores each.
We integrate the likelihood from $ f _{ \text{min}} = 20 $ Hz and generate waveforms from an orbit-averaged $ (2,2) $-mode frequency $ \langle f _{ \text{start}} \rangle = 10 $ Hz for GW150914; for GW190521 we use 11 and 5.5 Hz, respectively, following Ref.~\cite{LIGOScientific:2020iuh}, and for the neutron-star--BH and binary neutron star signals~\cite{LIGOScientific:2021qlt, LIGOScientific:2017vwq} we start both at the likelihood frequency (20 and 23 Hz, respectively), following the corresponding LVK analyses.
For GW170817 we also fix the sky location to the host galaxy NGC~4993 and restrict the spin magnitudes to $ | \bm \chi_i | \leq 0.05 $~\cite{LIGOScientific:2017vwq}.

%%%%%%%%%%%%%%%%%%%%%%%%%%%
%%%%%%%%%%%%%%%%%%%%%%%%%%%
%%%%%%%%%%%%%%%%%%%%%%%%%%%
%%%%%%%%%%%%%%%%%%%%%%%%%%%

%\appendix

\onecolumngrid

\section{Supplemental Material}
\label{sec:supplemental}

\twocolumngrid

%%%%%%%%%%%%%%%%%%%%%%%%%%%
%%%%%%%%%%%%%%%%%%%%%%%%%%%

\paragraph{Precession evolution equations}
\phantomsection\label{sm:precession}

The precession formulas $ \big \{ \bm{\Omega}_{S_{\! i}}^{ \text{qc}}, \dot{\bm l}_{ \text{N}} ^{ \text{\s qc}}, \bm{L} ^{ \text{qc}} \big \} $ employed in the equations of motion \eqref{eq:EOM} are derived through an orbit-average procedure for QC orbits up to 4PN order, with spin-orbit and spin-spin contributions at next-to-next-to-leading order~\cite{Khalilv5}.
The functions $ \bm{\Omega}_{S_{\! i}}^{ \text{qc}} $ and $ \dot{\bm l}_{ \text{N}} ^{ \text{qc}} $ are used in \texttt{SEOBNRv5PHM} and are given in Eqs.~(66) and (71) of Ref.~\cite{Khalilv5}, while $ \bm{L} ^{ \text{qc}}$ is derived here.

In \texttt{SEOBNRv5PHM}, the orbital angular momentum is obtained from a 3.5PN algebraic expression $ \bm{L} ^{ \text{qc}} (\bm{S}_{ 1}, \bm{S}_{ 2}, \bm{l}_{ \text{N}}, v) $ (Eqs.~(65) of Ref.~\cite{Khalilv5}), with $ v = (M \Omega)^{1/3} $.
Such a prescription poses no problem, since its precession equations are evolved \emph{before} the EOB equations of motion.
In our case, however, Eqs.~\eqref{eq:EOM} are solved as a fully coupled system to capture eccentric effects on the spin dynamics, and hence the expression $ \bm{L} ^{ \text{qc}} (\Omega) $ becomes implicit: the frequency follows from the Hamiltonian as $ \Omega = \partial H _{ \text{EOB}} ^{ \text{pprec}} / \partial p_\phi $, and the Hamiltonian itself depends on the projections of the spins onto $ \bm{L} $.
We remove the implicit dependence by trading $ v $ for the separation $ r $.
For QC orbits, the frequency is a function of a single orbital parameter, so the PN relation between $ v $ and $ r $ (Eqs.~(A2) of Ref.~\cite{Khalilv5}) can be inverted perturbatively, order by order, to give $ v(r) $.
Substituting $ v(r) $ into $ \bm{L} ^{ \text{qc}} $ yields
\begin{widetext}
\begin{subequations}
\label{eq:L_r}
\begin{align}
\bm{L} ^{ \text{qc}}
&\equiv
\bar{\boldsymbol{L}}_{S^0}+ \bar{\boldsymbol{L}}_\text{SO} + \bar{\boldsymbol{L}}_{S_{\!1}S_{\!2}} + \bar{\boldsymbol{L}}_{S^2},
\\[4pt]
\bar{\boldsymbol{L}}_{S^0} &=
\mu M \, \boldsymbol{l}_{\rm N} \, \Bigg\{
\sqrt{\frac{r}{M}}+\frac{3}{2}\,\sqrt{\frac{M}{r}}
+ \frac{M^{3/2}}{r^{3/2}} \left(\frac{27}{8}-\frac{3\nu}{2}\right)
+ \frac{M^{5/2}}{r^{5/2}} \left[\frac{135}{16}+\left(\frac{41\pi^{2}}{32}-\frac{433}{12}\right)\nu\right]
\nonumber\\%[4pt]
&\quad + \frac{M^{7/2}}{r^{7/2}} \left[\frac{2835}{128}
+ \nu\left(-\frac{3029}{120}-\frac{3503\pi^{2}}{2048}
-32\gamma_{E}-64\ln 2+16\ln \left( \frac{r}{M}\right)\right)
+ \left(\frac{539}{12}-\frac{205\pi^{2}}{128}\right)\nu^{2}\right]
\Bigg\},
\\[4pt]
\bar{\boldsymbol{L}}_\text{SO} &=
\frac{M}{r}\left[
\left(-\frac{3X_2}{4}-\frac{\nu}{4}\right)\boldsymbol{S}_1
+\left(-\frac{3X_2}{2}-\frac{\nu}{2}\right)\left(\boldsymbol{l}_{\rm N}\cdot\boldsymbol{S}_1\right)\boldsymbol{l}_{\rm N}
\right]
+\frac{M^2}{r^{2}}
\Bigg\{
\left[-\frac{9X_2}{16}+\left(\frac{3X_2}{8}-\frac{15}{16}\right)\nu-\frac{\nu^{2}}{16}\right]\boldsymbol{S}_1
\nonumber\\%[4pt]
&\qquad
+\left[-\frac{75X_2}{32}+\left(\frac{27X_2}{16}-\frac{117}{32}\right)\nu+\frac{3\nu^{2}}{32}\right]
\left(\boldsymbol{l}_{\rm N}\cdot\boldsymbol{S}_1\right)\boldsymbol{l}_{\rm N}
\Bigg\}
+\frac{M^3}{r^{3}}
\Bigg\{
\left[
-\frac{27X_2}{32}+\left(\frac{87X_2}{16}-\frac{81}{32}\right)\nu
+\left(\frac{45}{32}+\frac{3X_2}{32}\right)\nu^{2}-\frac{\nu^{3}}{32}\Bigg] \, \boldsymbol{S}_1
\right.
\nonumber\\%[4pt]
&\qquad
+\Bigg[-\frac{1769X_2}{256}+\left(\frac{385X_2}{16}-\frac{3415}{256}\right)\nu
+\left(\frac{915}{128}-\frac{11X_2}{256}\right)\nu^{2} +\frac{67\nu^{3}}{768}\Bigg]
\left(\boldsymbol{l}_{\rm N}\cdot\boldsymbol{S}_1\right)\boldsymbol{l}_{\rm N}
\Bigg\}
+1\leftrightarrow 2,
\\[4pt]
\bar{\boldsymbol{L}}_{S_{\!1}S_{\!2}} &=
\frac{M^{3/2}}{\mu M \, r^{3/2}}\Bigg\{
\frac{\nu}{2}\left(\boldsymbol{l}_{\rm N}\cdot\boldsymbol{S}_1\right)\boldsymbol{S}_2
+\frac{\nu}{2}\left(\boldsymbol{l}_{\rm N}\cdot\boldsymbol{S}_2\right)\boldsymbol{S}_1
+\left[
\left(-\frac{1}{2}+2\nu\right)
\left(\boldsymbol{l}_{\rm N}\cdot\boldsymbol{S}_1\right)\left(\boldsymbol{l}_{\rm N}\cdot\boldsymbol{S}_2\right)
+\left(\frac{1}{2}-\nu\right)\left(\boldsymbol{S}_1\cdot\boldsymbol{S}_2\right)
\right]\boldsymbol{l}_{\rm N}
\Bigg\}
\nonumber\\%[4pt]
&\quad +\frac{M^{5/2}}{\mu M \, r^{5/2}}\Bigg\{
\frac{7\nu}{4}\left(\boldsymbol{l}_{\rm N}\cdot\boldsymbol{S}_1\right)\boldsymbol{S}_2
+\frac{7\nu}{4}\left(\boldsymbol{l}_{\rm N}\cdot\boldsymbol{S}_2\right)\boldsymbol{S}_1
+\Bigg[
\left(\frac{221}{96}+\frac{25\nu}{36}+\frac{221\nu^{2}}{144}\right)
\left(\boldsymbol{l}_{\rm N}\cdot\boldsymbol{S}_1\right)\left(\boldsymbol{l}_{\rm N}\cdot\boldsymbol{S}_2\right)
\nonumber\\%[4pt]
&\qquad \quad
+\left(-\frac{31}{32}+\frac{\nu}{12}+\frac{\nu^{2}}{6}\right)
\left(\boldsymbol{S}_1\cdot\boldsymbol{S}_2\right)
\Bigg] \, \boldsymbol{l}_{\rm N}
\Bigg\}
+\frac{M^{7/2}}{\mu M \, r^{7/2}}\Bigg\{
\left[\left(\frac{21}{8}+\frac{15X_2}{16}\right)\nu
-\left(\frac{339}{64}+\frac{9X_2}{64}\right)\nu^{2}
-\frac{45\nu^{3}}{64}\right]\left(\boldsymbol{l}_{\rm N}\cdot\boldsymbol{S}_1\right)\boldsymbol{S}_2
\nonumber\\%[4pt]
&\qquad 
+\left[\left(\frac{57}{16}-\frac{15X_2}{16}\right)\nu
+\left(-\frac{87}{16}+\frac{9X_2}{64}\right)\nu^{2}
-\frac{45\nu^{3}}{64}\right]
\left(\boldsymbol{l}_{\rm N}\cdot\boldsymbol{S}_2\right)\boldsymbol{S}_1
+\Bigg[
\left(-\frac{639}{128}+\frac{19\nu}{128}
-\frac{4801\nu^{2}}{576}+\frac{\nu^{3}}{9}\right)
\left(\boldsymbol{S}_1\cdot\boldsymbol{S}_2\right)
\nonumber\\%[4pt]
&\qquad\quad
+\left(\frac{287}{128}+\frac{13201\nu}{1152}
-\frac{17507\nu^{2}}{3456}+\frac{1057\nu^{3}}{1728}\right)
\left(\boldsymbol{l}_{\rm N}\cdot\boldsymbol{S}_1\right)\left(\boldsymbol{l}_{\rm N}\cdot\boldsymbol{S}_2\right)
\Bigg]\, \boldsymbol{l}_{\rm N}
\Bigg\},
\\[4pt]
\bar{\boldsymbol{L}}_{S^2} &=
\frac{M^{3/2}}{\mu M \, r^{3/2}}\Bigg\{
\left(\frac{X_2}{2}-\frac{\nu}{2}\right)\left(\boldsymbol{l}_{\rm N}\cdot\boldsymbol{S}_1\right)\boldsymbol{S}_1
+\left[
\left(\frac{3X_2}{4}-\frac{3\nu}{4}\right)\left(\boldsymbol{l}_{\rm N}\cdot\boldsymbol{S}_1\right)^{2}
+\left(-\frac{X_2}{4}+\frac{\nu}{4}\right)S_1^{2}
\right]\boldsymbol{l}_{\rm N}
\Bigg\}
\nonumber\\%[4pt]
&\quad +\frac{M^{5/2}}{\mu M \, r^{5/2}}\Bigg\{
\left(\frac{7X_2}{4}-\frac{7\nu}{4}\right)\left(\boldsymbol{l}_{\rm N}\cdot\boldsymbol{S}_1\right)\boldsymbol{S}_1
+\Bigg[
\left(-\frac{49X_2}{32}+\left(\frac{49}{32}+\frac{81X_2}{16}\right)\nu
+\frac{35\nu^{2}}{32}\right)\left(\boldsymbol{l}_{\rm N}\cdot\boldsymbol{S}_1\right)^{2}
\nonumber\\%[4pt]
&\qquad
+\left(\frac{5X_2}{16}+\left(-\frac{5}{16}-\frac{11X_2}{8}\right)\nu
-\frac{5\nu^{2}}{16}\right)S_1^{2}
\Bigg] \, \boldsymbol{l}_{\rm N}
\Bigg\}
+\frac{M^{7/2}}{\mu M \,  r^{7/2}}\Bigg\{
\Bigg[\frac{87X_2}{64}+\left(-\frac{87}{64}-\frac{3X_2}{2}\right)\nu
+\left(\frac{405}{64}-\frac{225X_2}{64}\right)\nu^{2}
\nonumber\\%[4pt]
&\qquad \quad
-\frac{45\nu^{3}}{64}\Bigg]\left(\boldsymbol{l}_{\rm N}\cdot\boldsymbol{S}_1\right)\boldsymbol{S}_1
+\Bigg[
\Bigg(-\frac{49X_2}{64}+\left(\frac{49}{64}+\frac{739X_2}{96}\right)\nu
+\left(\frac{3043}{192}-\frac{103X_2}{16}\right)\nu^{2}
-\frac{3\nu^{3}}{16}\Bigg)\left(\boldsymbol{l}_{\rm N}\cdot\boldsymbol{S}_1\right)^{2}
\nonumber\\%[4pt]
&\qquad \quad
+\Bigg(-\frac{361X_2}{128}+\left(\frac{361}{128}-\frac{79X_2}{24}\right)\nu
+\left(\frac{295}{384}+\frac{287X_2}{128}\right)\nu^{2}
+\frac{51\nu^{3}}{128}\Bigg) \, S_1^{2}
\Bigg] \, \boldsymbol{l}_{\rm N}
\Bigg\}
+1\leftrightarrow 2,
\end{align}
\end{subequations}
\end{widetext}
where $ X_i = m_i / M $ ($ i \in \{1, 2\} $).
Since $ r $ is one of the variables evolved in Eqs.~\eqref{eq:EOM}, the circular dependence is broken: $ \bm{L} $ is evaluated explicitly at each step.

%%%%%%%%%%%%%%%%%%%%%%%%%%%
%%%%%%%%%%%%%%%%%%%%%%%%%%%

\paragraph{Additional NR-waveform comparisons}
\phantomsection\label{sm:nr}

\begin{figure*}
\hspace{-6pt}
\includegraphics[width=0.95\linewidth]{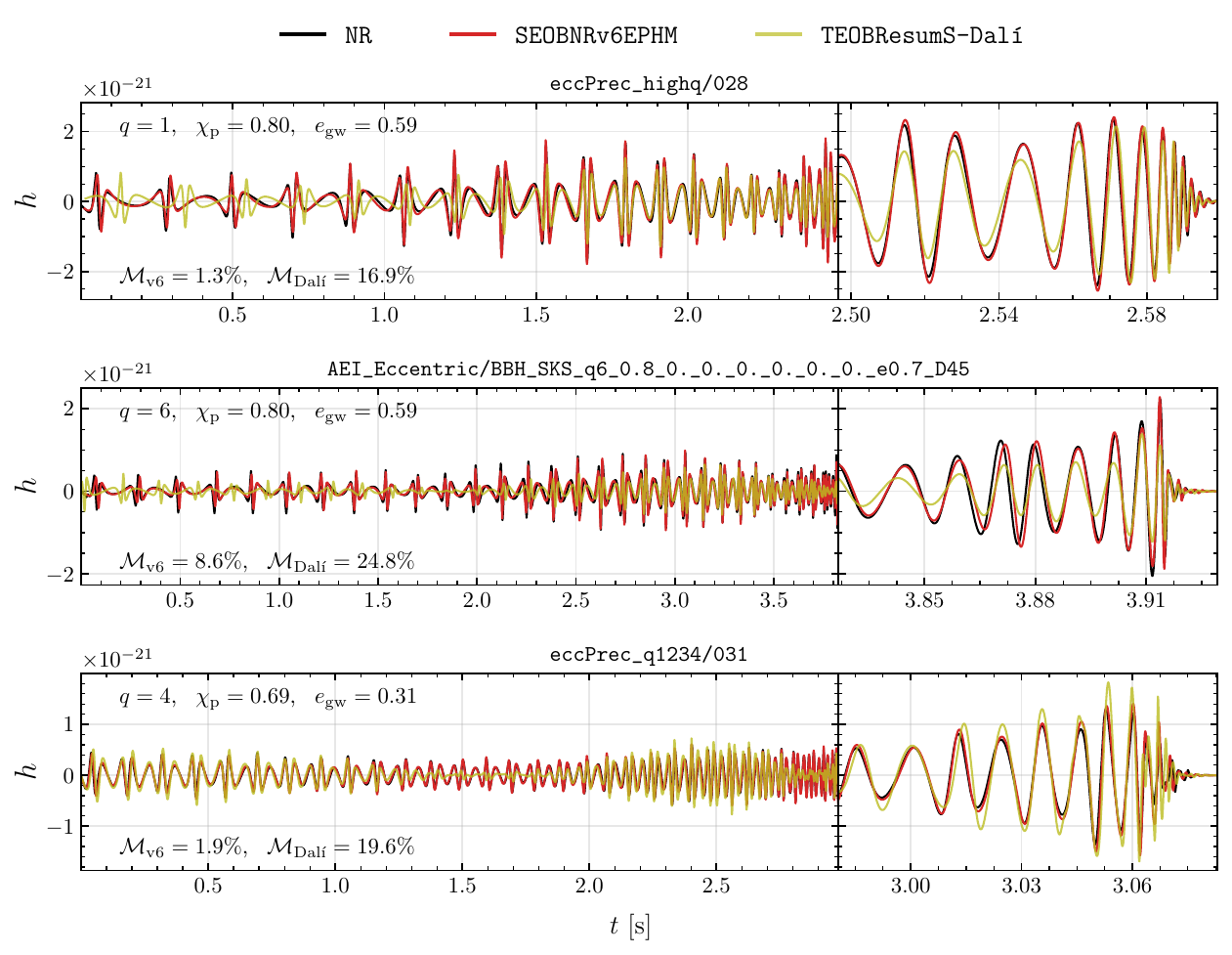}
\vspace{-5pt}
\caption{
GW strain $ h $ from NR (black), \texttt{SEOBNRv6EPHM} (red), and \texttt{TEOBResumS-Dal\'i} (yellow) for three representative eccentric, spin-precessing configurations, with simulation name, mass ratio $ q $, effective precession spin $ \chi_{\text{p}} $, and initial GW eccentricity $ e_{\text{gw}} $ annotated in each row.
The waveforms correspond to BBHs with total mass $ M = 40 \, \solarmass $ and distance $ d_{L} = 100 $\,Mpc, observed at inclination $ \iota_{\text{s}} = \pi/3 $, azimuthal angle $ \varphi _{ \text{s}} = \pi / 2 $, and polarization angle $ \kappa_{ \text{s}} = \pi / 2 $, with the template parameters set by the mismatch optimization.
Each row quotes the corresponding mismatches of \texttt{SEOBNRv6EPHM} ($ \mathcal{M}_{\text{v6}} $) and \texttt{TEOBResumS-Dal\'i} ($ \mathcal{M}_{\text{Dal\'i}} $).
Right panels zoom into the merger-ringdown.
}
\label{fig:waveform_comparisons}
\end{figure*}
\begin{figure*}
\hspace{-5pt}
\vspace{8pt}
\includegraphics[width=0.78\linewidth]{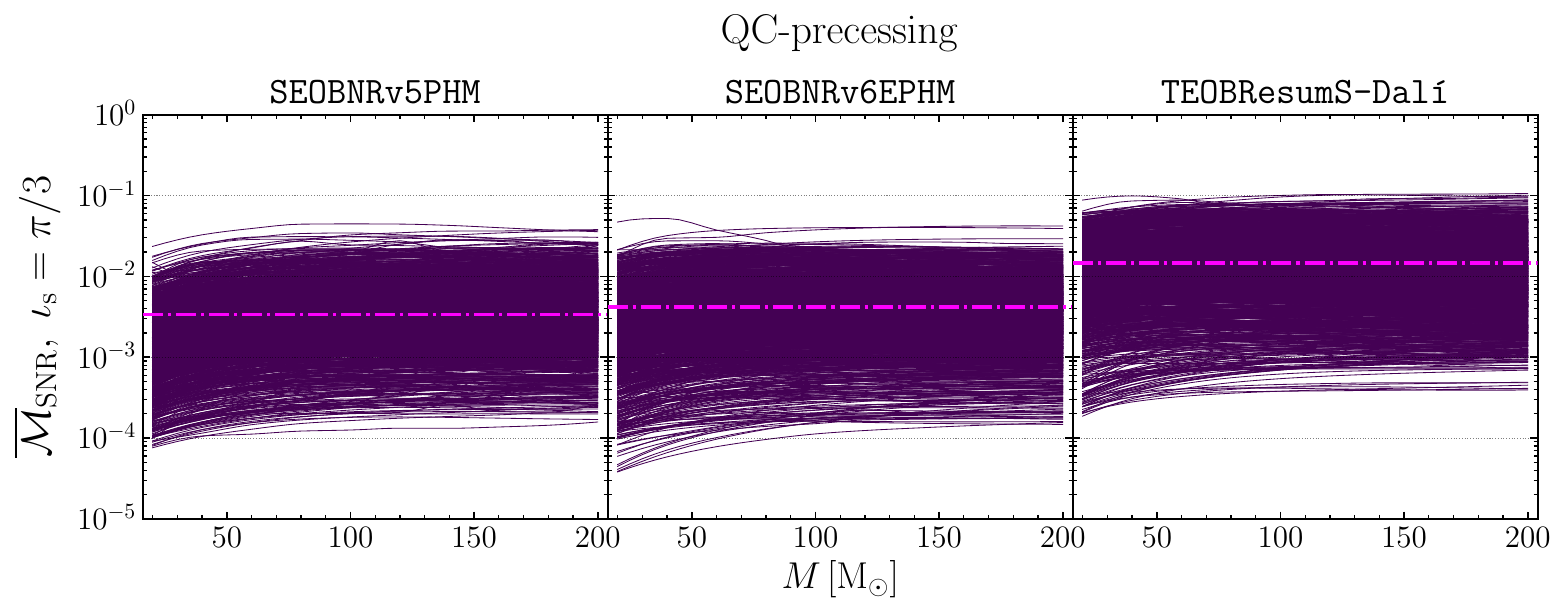}
\vspace{8pt}
\includegraphics[width=0.575\linewidth]{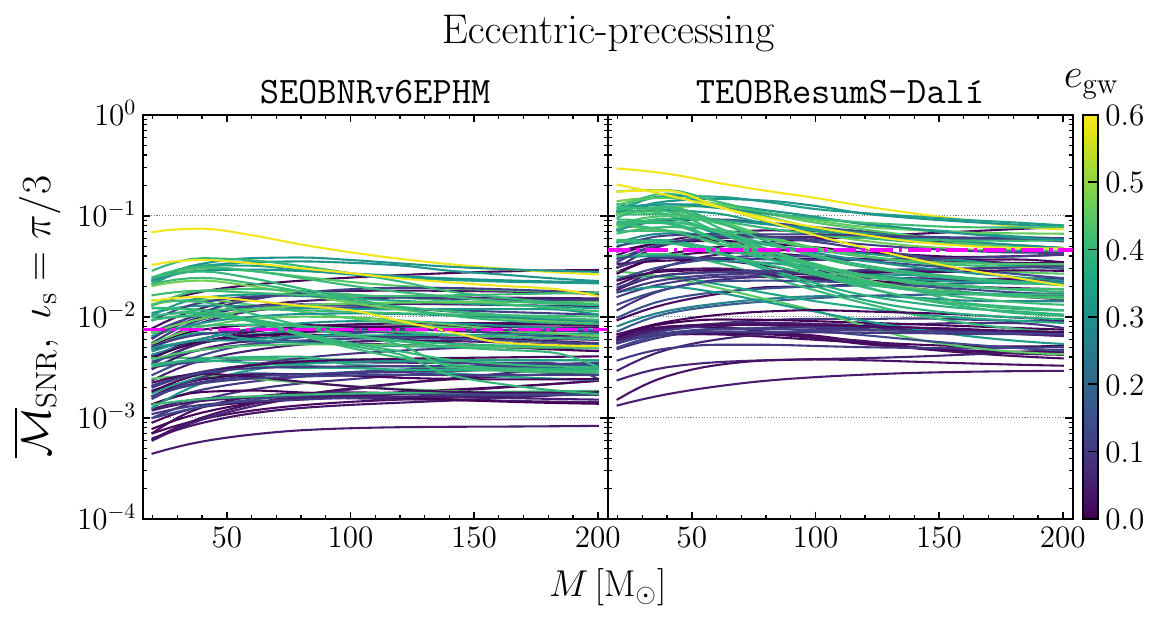}
\hspace{80pt}
\includegraphics[width=0.315\linewidth]{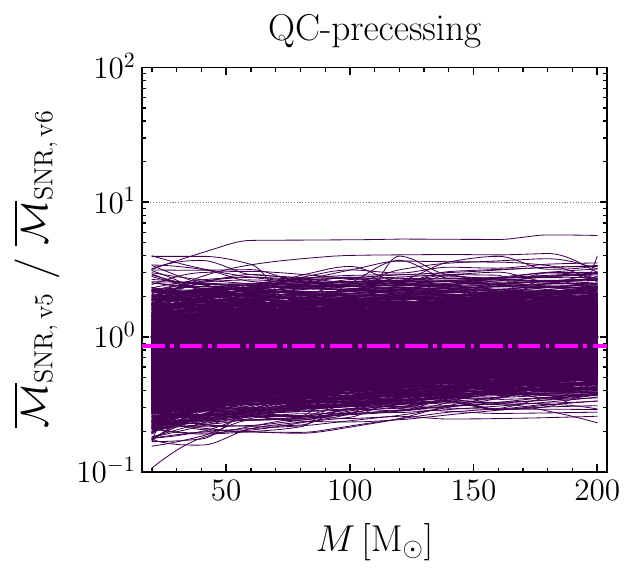}
\hspace{1pt}
\includegraphics[width=0.315\linewidth]{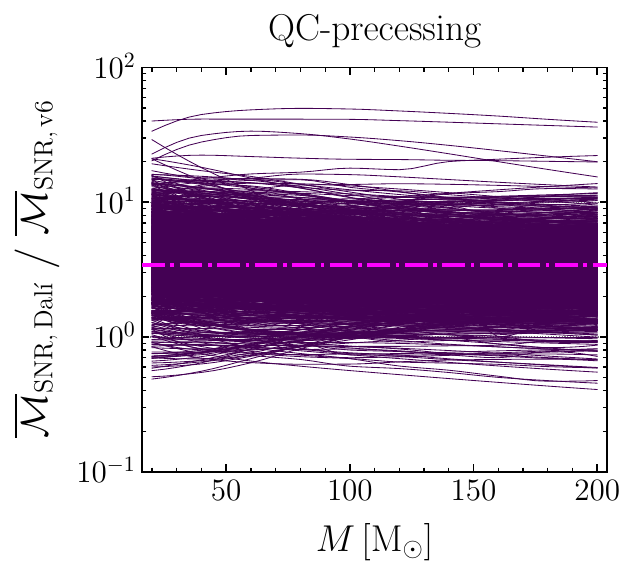}
\hspace{1pt}
\includegraphics[width=0.344\linewidth]{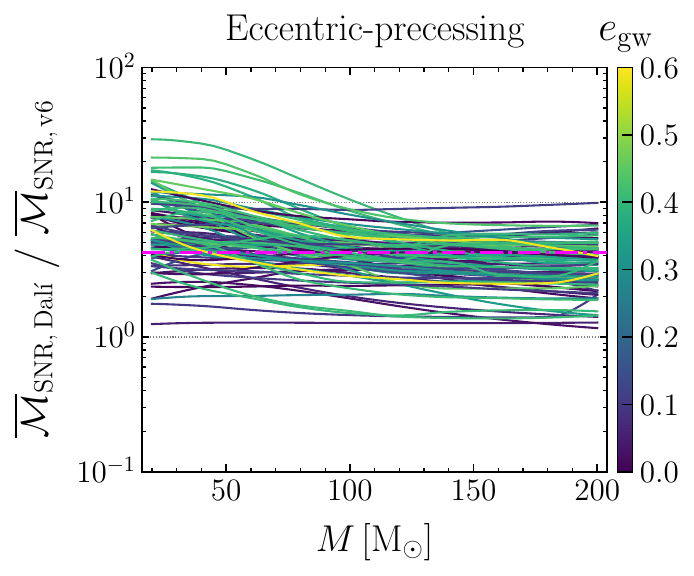}
\vspace{-5pt}
\caption{
Waveform mismatches for different models against 1437 QC and 87 eccentric \texttt{SXS} NR waveforms of spin-precessing BBHs, computed over the total mass range $ [20, \s 200] \, \solarmass $ at inclination $ \iota _{ \text{s}} = \pi/3 $.
The first and second rows show the SNR-weighted mismatch of each model against the QC and eccentric simulations, respectively.
The third row shows the ratio of the \texttt{SEOBNRv5PHM} (left) or \texttt{TEOBResumS-Dal\'i} (middle and right) mismatch to that of \texttt{SEOBNRv6EPHM}, for the QC (left and middle) and eccentric (right) cases; values above unity indicate that \texttt{SEOBNRv6EPHM} is more accurate.
Curve color indicates the initial GW eccentricity $ e_{\text{gw}} $ of each NR waveform.
Magenta dot-dashed lines mark the median of the per-simulation maximum mismatches (first two rows) and of the per-simulation median ratios (third row).
}
\label{fig:mms_v5_v6_dali}
\end{figure*}
\begin{figure*}
\hspace{-6pt}
\includegraphics[width=0.485\linewidth]{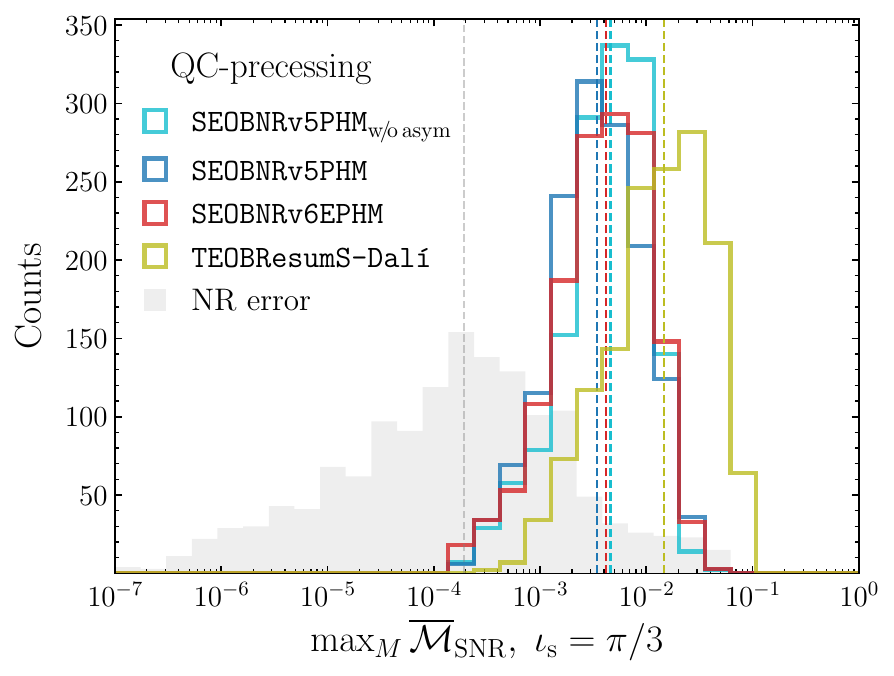}
\hspace{5pt}
\includegraphics[width=0.478\linewidth]{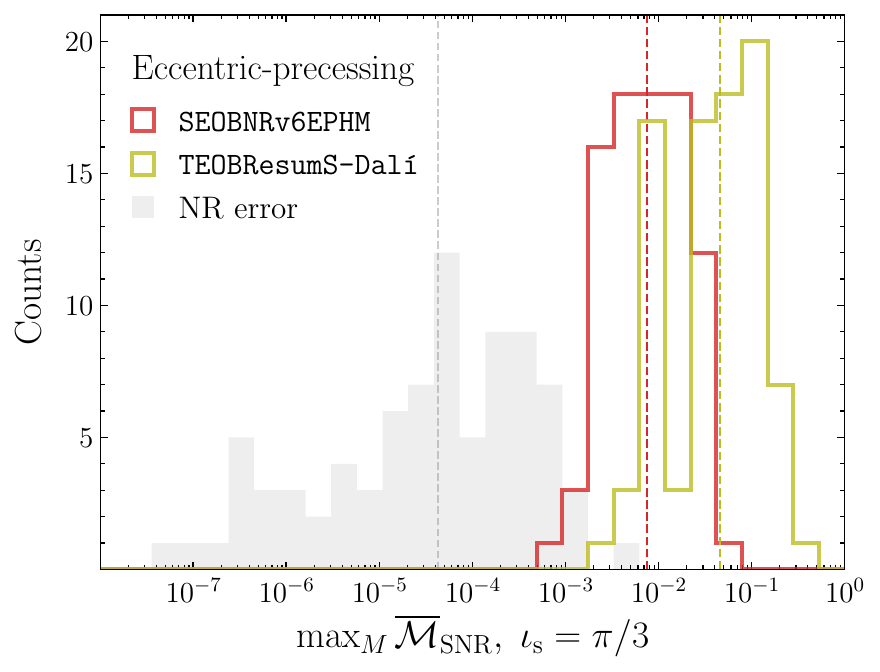}
\vspace{-5pt}
\caption{
Distributions of the maximum SNR-weighted mismatches over the total mass range $ [20, \s 200] \, \solarmass $ at inclination $ \iota _{ \text{s}} = \pi/3 $ for different models against 1437 QC (left) and 87 eccentric (right) spin-precessing \texttt{SXS} NR waveforms.
\texttt{SEOBNRv5PHM} includes mode asymmetries, while \texttt{SEOBNRv5PHM}$_\text{w/o asym}$ omits them.
The NR error histograms (gray) are obtained from the waveform mismatches between the highest and second-highest resolutions.
Dashed lines mark the median of each distribution.
}
\label{fig:hist_mms}
\end{figure*}
\begin{figure*}
\hspace{-5pt}
%\vspace{2pt}
\includegraphics[width=0.95\linewidth]{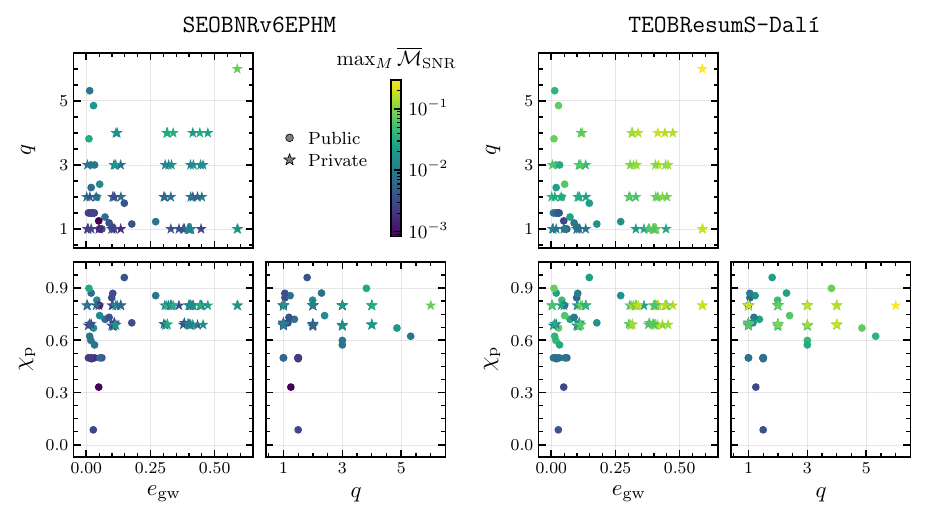}
\vspace{-15pt}
\caption{
Distribution of the 87 eccentric, spin-precessing \texttt{SXS} NR simulations ($ 62 $ private + $ 25 $ public) used in this work, characterized by their mass ratio $ q $, effective precession spin $ \chi _{ \text{p}} $, and initial GW eccentricity $ e _{ \text{gw}} $, color-coded by the maximum SNR-weighted mismatch over $ M \in [20, \s 200] \, \solarmass $, for \texttt{SEOBNRv6EPHM} (left) and \texttt{TEOBResumS-Dal\'i} (right).
}
\label{fig:mismatches_corner}
\end{figure*}

We provide complementary information on the mismatches of \texttt{SEOBNRv5PHM}, \texttt{SEOBNRv6EPHM}, and \texttt{TEOBResumS-Dal\'i} against 1437 QC and 87 eccentric \texttt{SXS} NR waveforms of spin-precessing BBHs, calculated over the total mass range $ [20, \s 200] \, \solarmass $ at inclination $ \iota _{ \text{s}} = \pi/3 $.
The \texttt{SEOBNRv5PHM} mismatches differ slightly from those in Refs.~\cite{RamosBuadesv5,Estelles:2025zah}, mainly because we consider NR waveforms with mode content up to $ \ell = 4 $ (rather than $ \ell = 5 $, since \texttt{SEOBNRv6EPHM} does not currently model the $ (5,5) $ mode) and use the latest \texttt{SXS} catalog~\cite{Scheel:2025jct}.

As proof of principle, we provide in Fig.~\ref{fig:waveform_comparisons} specific examples from our eccentric, spin-precessing waveform comparisons.
We plot the strain $ h $ as a function of time (Eq.~\eqref{eq:h}, with $ \mathcal A = 1 $) for three NR waveforms, together with the best-fit predictions from \texttt{SEOBNRv6EPHM} and \texttt{TEOBResumS-Dal\'i}.
We consider a highly eccentric simulation ($ e _{ \text{gw}} = 0.59 $) with mass ratio $ q = 1 $ (first row, $ \max_M \overline{\mathcal{M}}_{\text{SNR}} = 1.6 \% $), and simulations for which \texttt{SEOBNRv6EPHM} has the highest (second row, $ \max_M \overline{\mathcal{M}}_{\text{SNR}} = 7.5 \% $) and second-highest (third row, $ \max_M \overline{\mathcal{M}}_{\text{SNR}} = 3.9 \% $) SNR-weighted mismatches over the considered mass range.
Each row shows the \emph{individual} mismatch $ \mathcal{M} $ for each model, corresponding to a binary with $ M = 40 \, \solarmass $ and source angles $ \varphi _{ \text{s}} = \kappa_{ \text{s}} = \pi / 2 $.
Despite the challenging configurations, we observe good agreement in amplitude and phase between NR and \texttt{SEOBNRv6EPHM} waveforms over their entire length, resulting in a percent-level mismatch.
In contrast, \texttt{TEOBResumS-Dal\'i} waveforms are in phase only for certain segments, and their overall amplitude does not match the NR predictions, resulting in much higher mismatches.
These outcomes are analogous to the results obtained in the aligned-spin limit (see Fig.~13 of Ref.~\cite{Gamboa:2026jht}).

Moving on to the full set of NR waveforms, Fig.~\ref{fig:mms_v5_v6_dali} extends Fig.~\ref{fig:mms_qc_ecc_v6} of the main text to all the considered models, showing the mismatches for the QC (first row) and eccentric (second row) comparisons, together with their ratio to those of \texttt{SEOBNRv6EPHM} (third row; in this format, a ratio above unity indicates that \texttt{SEOBNRv6EPHM} is more accurate).
Figure~\ref{fig:hist_mms} then summarizes the information with histograms of the maximum mismatches over the considered mass range, with an estimation of the numerical error of the NR waveforms;
for the QC case, we show the results when including (\texttt{SEOBNRv5PHM}) or not including (\texttt{SEOBNRv5PHM}$_\text{w/o asym}$) mode asymmetries.

From these results, we conclude that \texttt{SEOBNRv5PHM} and \texttt{SEOBNRv6EPHM} share the same accuracy level in the QC-orbit limit, and that \texttt{SEOBNRv6EPHM} is a median factor of $ \sim 4 $ more accurate than \texttt{TEOBResumS-Dal\'i}.
For QC waveforms, the ratio of mismatches between \texttt{SEOBNRv5PHM} and \texttt{SEOBNRv6EPHM} stays, on average, close to the unity level (despite the latter not including mode asymmetries; if one omits mode asymmetries in \texttt{SEOBNRv5PHM}, the median accuracy of \texttt{SEOBNRv6EPHM} is slightly better, as seen in Fig.~\ref{fig:hist_mms}).
This is expected since both models share a similar baseline (the same result is observed between the aligned-spin models \texttt{SEOBNRv5HM} and \texttt{SEOBNRv6EHM}~\cite{Gamboa:2026jht}).
For the majority of QC and all eccentric waveforms, \texttt{SEOBNRv6EPHM} outperforms \texttt{TEOBResumS-Dal\'i}, in some cases by more than an order of magnitude, and by comparable factors in the two cases, with a median improvement factor of $ \sim 4 $ (see last row of Fig.~\ref{fig:mms_v5_v6_dali}).

The distribution of mismatches across the eccentric, spin-precessing parameter space, however, is nontrivial.
Figure~\ref{fig:mismatches_corner} shows the accuracy of \texttt{SEOBNRv6EPHM} and \texttt{TEOBResumS-Dal\'i} across the parameter space of the 87 eccentric NR waveforms, characterized by their GW eccentricity $ e _{ \text{gw}} $, effective precession spin $ \chi _{ \text{p}} $, and mass ratio $ q $ (all obtained at the NR relaxation time).
Each marker represents a NR simulation color-coded by the maximum mismatch over the considered total-mass interval.
As expected, high mismatches appear at high mass ratios, large eccentricities, and for strongly precessing systems.

While it is true that increasing eccentricity results in more challenging configurations, the $ q $ vs $ e _{ \text{gw}} $ corner plot of \texttt{SEOBNRv6EPHM} (left panel of Fig.~\ref{fig:mismatches_corner}) shows that the major accuracy loss occurs as the mass ratio increases (markers become more yellow as $ q $ grows).
The same situation occurs with state-of-the-art QC waveform models~\cite{Mahapatra:2026wsp} and reflects the current inadequacy of comparable-mass models in dealing with large-mass-ratio systems.
For the moment, as mentioned in the main text, we emphasize that systems with $ q \leq 4$ keep mismatches below $ 4 \% $, regardless of the eccentricity and spin-precession values (for our considered set of NR simulations).
Nevertheless, even for the highly eccentric and highly precessing system with the largest mass ratio ($ e _{ \text{gw}} = 0.59 $, $ \chi _{ \text{p}} = 0.8 $, $ q=6 $), \texttt{SEOBNRv6EPHM} captures the correct phenomenology of the NR waveform, as shown in Fig.~\ref{fig:waveform_comparisons}.

We note that \texttt{TEOBResumS-Dal\'i} has the same accuracy challenge for large mass ratios \emph{and} increasing eccentricities:
the $ q $ vs $ e _{ \text{gw}} $ corner plot of \texttt{TEOBResumS-Dal\'i} (right panel of Fig.~\ref{fig:mismatches_corner}) shows a noticeable accuracy loss as either $ q $ or $ e _{ \text{gw}} $ grow.
This means that their eccentric sector is not sufficiently well modeled, as already found in the aligned-spin limit~\cite{Gamboa:2026jht}.
In particular, the dephasing of the \texttt{TEOBResumS-Dal\'i} best-fit waveforms with respect to NR in Fig.~\ref{fig:waveform_comparisons} suggests that its RR-force prescription may benefit from further refinement.

\begin{table*}
\caption{
Parameters inferred from GW200129 with the eccentric, spin-precessing model \texttt{SEOBNRv6EPHM}, for six treatments of the LIGO-Livingston glitch and two eccentricity priors.
Bayes factors are quoted against the model's QC ($e=0$) limit, the QC precessing model \texttt{NRSur7dq4}, and the model's aligned-spin ($\chi_\text{p}=0$) limit.
Model eccentricity $e_{10\,\text{Hz}}$, GW eccentricity $e_{\text{gw},10\,\text{Hz}}$, and $\chi_\text{p}$ are quoted as medians with 90\% credible intervals, at a $10$~Hz reference frequency.
Green (red) shading denotes support for (against) the eccentric hypothesis in the two $e$-columns and for (against) spin precession in the $\chi _{ \text{p}}=0$ column.
The nested-sampling uncertainty on every quoted Bayes factor is $\simeq 0.1$.
}
\input{tab/GW200129_table}
\label{tab:gw200129_table}
\end{table*}

As in the QC case, the accuracy of waveform approximants has not yet reached the numerical error of the NR simulations (see Fig.~\ref{fig:hist_mms}).
There is thus significant room for improvement through additional analytical information (e.g., eccentric terms in the spin and angular-momentum evolution equations), calibration to NR simulations, and refined phenomenological prescriptions for the most challenging portions of the waveforms (periastron passages and the plunge).
These developments, and further model validation, will require more eccentric, spin-precessing NR simulations.

%%%%%%%%%%%%%%%%%%%%%%%%%%%
%%%%%%%%%%%%%%%%%%%%%%%%%%%

\paragraph{Additional details of the GW200129 analysis}
\phantomsection\label{sm:gw200129}

As summarized in the main text, the analysis of GW200129 is complicated by a glitch overlapping the signal in LIGO-Livingston.
Glitch mitigations based on auxiliary witness channels---\texttt{gwsubtract}~\cite{LIGOScientific:2021djp} and NLSUB~\cite{Macas:2023wiw}---support precession~\cite{Hannam:2021pit}.
Ref.~\cite{Payne:2022spz} instead modeled the glitch jointly with the GW signal from strain data alone, using \texttt{BayesWave}~\cite{Cornish:2014kda, Chatziioannou:2021ezd} with sine-Gaussian wavelets and the QC aligned-spin model \texttt{IMRPhenomD}~\cite{Khan:2015jqa}, and found weaker support for precession when the glitch-subtracted data were reanalyzed with \texttt{NRSur7dq4}~\cite{Varma:2019csw}.
More recent joint analyses built on \texttt{Bilby}~\cite{Ashton:2018jfp, Romero-Shaw:2020owr}, using \texttt{NRSur7dq4} directly, reach differing conclusions depending on the glitch model: Ref.~\cite{Hoy:2026dkr} finds the precession measurement robust, whereas Ref.~\cite{Cheung:2026myt} finds no strong evidence.

Eccentric, aligned-spin analyses of GW200129 found varying support for eccentricity depending on the glitch treatment~\cite{Gupte:2024jfe,Planas:2025plq}, though no joint analysis of the GW signal and glitch has yet been performed with an eccentric model.
Since the \texttt{BayesWave} draws we adopt were obtained assuming a QC, aligned-spin signal~\cite{Payne:2022spz}, part of the eccentric power may have been absorbed into the glitch model, and the corresponding Bayes factors should be read as conservative.
More generally, subtracting a glitch leaves residual power that can bias the inferred parameters, and jointly inferring the glitch and GW signal typically recovers them more faithfully~\cite{Udall:2025bts}.
The impact of waveform mismodeling---such as neglected eccentricity---on these joint analyses has not yet been quantified; we will address it in upcoming work through a joint eccentric, spin-precessing analysis of the signal and glitch.

All GW200129 analyses use $ 2000 $ live points (\texttt{nlive}), twice the baseline value adopted for other events, to improve sampler convergence and the accuracy of the evidence estimates;
the remaining settings are as described in the \EndMatter[em:pe_settings].
We integrate the likelihood from $ f _{ \text{min}} = 20 $ Hz and sample the spin and eccentric parameters at an orbit-averaged $ (2,2) $-mode frequency $ \langle f _{ \text{start}} \rangle = 10 $ Hz, from which the waveforms are extended backward in time by $10000\, M$.
This extension captures signal power from early periastron passages (whose instantaneous frequencies exceed the orbit-averaged value) and can tighten the eccentricity constraints~\cite{Pompili:2026yxq}.

Table~\ref{tab:gw200129_table} collects the results for the six glitch treatments and the two eccentricity priors.
The eccentric hypothesis is favored over both the QC limit of \texttt{SEOBNRv6EPHM} and \texttt{NRSur7dq4} in every case, with $ \log_{10}\mathcal{B} $ largest for the unmitigated and auxiliary-channel data, and smallest for the \texttt{BayesWave} draws, mirroring the ordering of the inferred eccentricity.
The model eccentricity $ e _{ 10 \, \text{Hz}} $ and the gauge-invariant $ e _{ \text{gw},\s 10 \, \text{Hz}} $~\cite{Shaikh:2023ypz,Shaikh:2025tae} agree to within the quoted precision.
Adopting a log-uniform prior lowers $ \log_{10}\mathcal{B} $ by $ \simeq 0.5 $--$0.8 $, but does not qualitatively change the inferred eccentricity, so the support for it is prior-sensitive in significance but robust.
By contrast, the Bayes factors for spin precession remain $\lesssim 0.5$ for all treatments, and $ \chi _{ \text{p}} $ is only weakly constrained.

\begin{table*}
\caption{
Median values and $90\%$ credible intervals of the source-frame total mass $M_\text{src}$, inverse mass ratio $1/q$, effective aligned spin $\chi_\text{eff}$, effective precessing spin $\chi_\text{p}$, and eccentricity $e$ inferred with \texttt{SEOBNRv6EPHM} for a set of GW events, together with the log Bayes factors $\log_{10}\mathcal{B}^{e\neq0}_{e=0}$ for the eccentric over the QC hypothesis and $\log_{10}\mathcal{B}^{\chi_\text{p}\neq0}_{\chi_\text{p}=0}$ for the precessing over the aligned-spin hypothesis. Events are ordered by decreasing $\log_{10}\mathcal{B}^{e\neq0}_{e=0}$;
green (red) shading denotes support for (against) each hypothesis, with independent color scales per column.
The nested-sampling uncertainty on $\log_{10}\mathcal{B}^{\chi_\text{p}\neq0}_{\chi_\text{p}=0}$ is $\simeq 0.1$ for all events, and is therefore not quoted per row.
The values of $e$ and $\chi_\text{p}$ are quoted at the orbit-averaged waveform starting frequency: $5.5$~Hz for GW190521, $20$~Hz for GW200105, $23$~Hz for GW170817, and $10$~Hz otherwise.
All analyses use a uniform prior on eccentricity, $ e \in [0, 0.8] $.
$^\ddag$~GW200129 is analyzed in detail in the main text (Fig.~\ref{fig:gw200129_eccentricity} and Table~\ref{tab:gw200129_table}); the row here uses the \texttt{gwsubtract} data and the uniform $e$ prior.
}
\input{tab/real_events_table}
\label{tab:real_events_table}
\end{table*}

We stress that these Bayes factors are not odds ratios.
The odds of the eccentric over the QC hypothesis, $ \mathcal{O} = \mathcal{R} _{ \text{e}}/\mathcal{R} _{ \text{qc}} \times \, \mathcal{B} $, also require the prior odds, i.e., the relative merger rates of eccentric $ \mathcal{R} _{ \text{e}} $ and QC $ \mathcal{R} _{ \text{qc}} $ binaries.
Ref.~\cite{Gupte:2024jfe} estimated these by combining the fractional rate of eccentric mergers expected in the galactic-nuclei, globular-cluster, young-star-cluster, and isolated-binary channels, obtaining a most probable prior odds $ \mathcal{R} _{ \text{e}}/\mathcal{R} _{ \text{qc}} \simeq 0.023 $; converting a Bayes factor into an odds ratio therefore lowers $ \log_{10}\mathcal{B} $ by $ \simeq 1.6 $.
This correction applies to the astrophysically agnostic uniform prior; the log-uniform prior instead concentrates its weight at small eccentricities and thus already encodes part of the expectation that few binaries are eccentric, thereby requiring a correspondingly milder prior-odds correction (a factor closer to unity).
Using the estimate of Ref.~\cite{Gupte:2024jfe} as an indicative value for the uniform prior, the support for eccentricity survives for the unmitigated, \texttt{gwsubtract} and NLSUB data ($ \log_{10}\mathcal{O} \simeq 1.9$--$3.7 $ against \texttt{NRSur7dq4}), but not uniformly for the \texttt{BayesWave} draws, which become marginal ($ \log_{10}\mathcal{O} \simeq 0.8 $ for draws B and C) or mildly disfavored (draw A).

%%%%%%%%%%%%%%%%%%%%%%%%%%%
%%%%%%%%%%%%%%%%%%%%%%%%%%%

\paragraph{Analysis of additional events}
\phantomsection\label{sm:events}

Beyond GW200129, we use \texttt{SEOBNRv6EPHM} to perform eccentric, spin-precessing PE analyses of ten additional GW events:
GW150914~\cite{LIGOScientific:2016aoc}, the first BBH detection, and the binary neutron star signal GW170817~\cite{LIGOScientific:2017vwq}, both expected to be QC and included as benchmarks;
GW190521~\cite{LIGOScientific:2020iuh}, a massive BBH event with early claims of eccentricity~\cite{Romero-Shaw:2020thy, Gayathri:2020coq, Gamba:2021gap};
the O3 BBH candidates GW190701\_203306 and GW200208\_222617~\cite{Gupte:2024jfe, Planas:2025jny};
the neutron-star--BH signal GW200105, for which several analyses report eccentricity support~\cite{Fei:2024ruj,Morras:2025xfu, Planas:2025plq, Kacanja:2025kpr, Jan:2025fps, Pompili:2026yxq};
the O4a candidates GW230712\_090405 and GW231223\_032836~\cite{Gupte:2026whi, Xu:2025ajj};
GW231123\_135430~\cite{LIGOScientific:2025rsn}, a signal from the most massive BBH observed, for which eccentricity has been reported against the QC aligned-spin but not the QC precessing hypothesis, depending on the waveform model~\cite{Jan:2025zcm, Malagon:2026uev, Xu:2025ajj}; and GW250114\_082203~\cite{LIGOScientific:2025rid}, the loudest BBH signal observed to date, included as a high-SNR benchmark.
The events GW150914, GW200105, GW231123\_135430, and GW250114\_082203 have been analyzed under the eccentric, spin-precessing hypothesis with the \texttt{TEOBResumS-Dal\'i} model in Refs.~\cite{Gamba:2025qfg,Jan:2025fps,Jan:2025zcm,Chandra:2025jfc}.

\begin{figure*}
\hspace{-5pt}
\includegraphics[width=0.8 \linewidth]{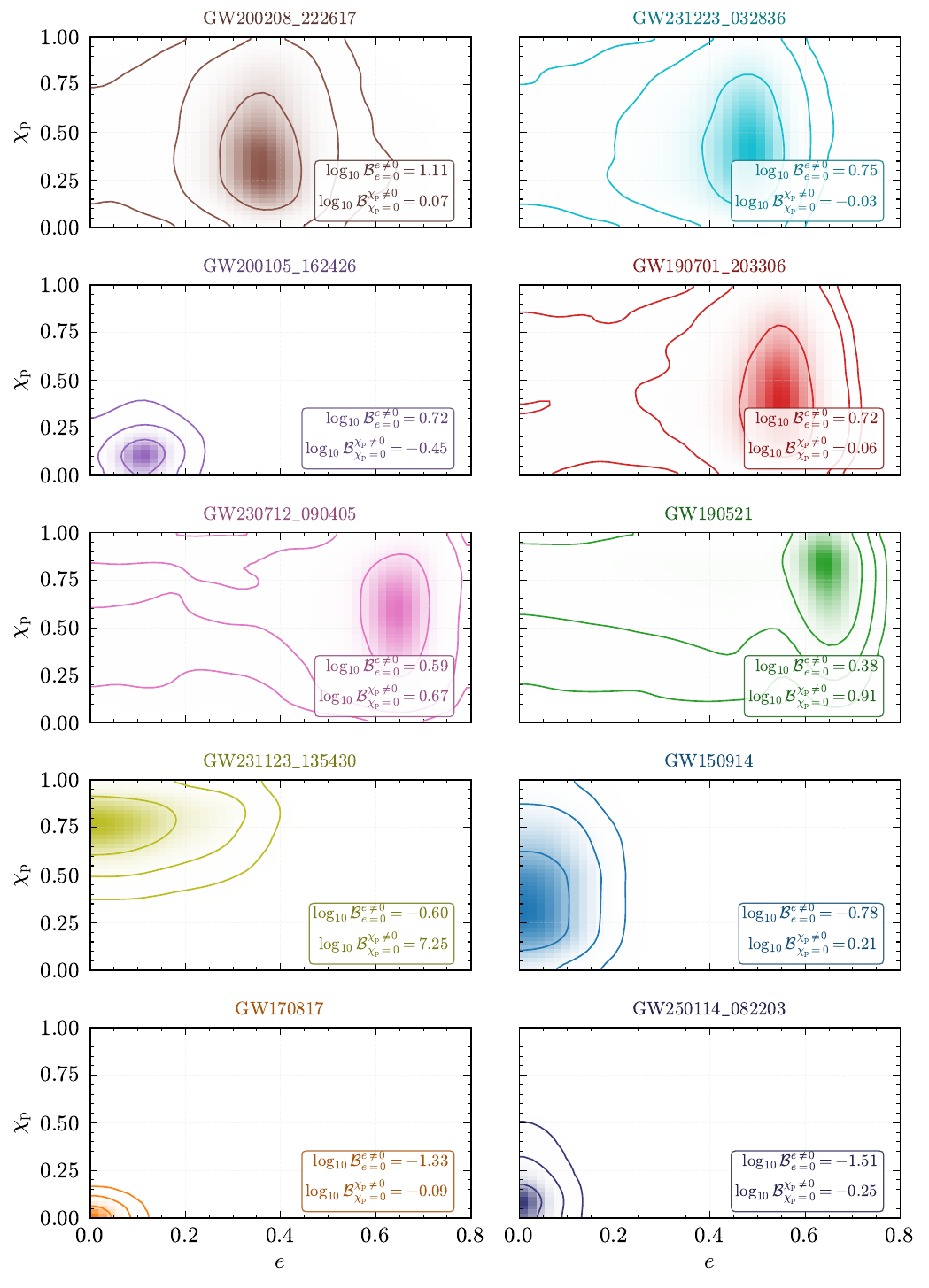}
\vspace{-5pt}
\caption{
Joint posteriors on the eccentricity $e$ and effective precessing spin $\chi_\text{p}$ inferred with \texttt{SEOBNRv6EPHM} for the events in Table~\ref{tab:real_events_table} (except GW200129, whose posterior is shown in Fig.~\ref{fig:gw200129_eccentricity} of the main text).
Each panel is annotated with the log Bayes factors $\log_{10}\mathcal{B}^{e\neq0}_{e=0}$ for eccentricity over the QC hypothesis and $\log_{10}\mathcal{B}^{\chi_\text{p}\neq0}_{\chi_\text{p}=0}$ for precession over the aligned-spin hypothesis. Contours enclose $50\%$, $90\%$, and $99\%$ credible regions.
The values of $e$ and $\chi_\text{p}$ are quoted at the orbit-averaged waveform starting frequency: $5.5$~Hz for GW190521, $20$~Hz for GW200105, $23$~Hz for GW170817, and $10$~Hz otherwise.
All analyses use a uniform prior on eccentricity, $ e \in [0, 0.8] $.
}
\label{fig:ecc_chip_corner_grid}
\end{figure*}

For each event, we use the strain data, PSDs, and calibration envelopes from GWOSC~\cite{LIGOScientific:2019lzm, KAGRA:2023pio}, adopting glitch-subtracted frames where data-quality issues are present.
We employ a uniform prior on eccentricity, $e \in [0, 0.8]$, and otherwise follow the data, prior, frequency, and sampler settings of the aligned-spin analysis of Ref.~\cite{Pompili:2026yxq}, except for the spin prior, which is taken to be isotropic in orientation and uniform in magnitude.
The eccentricity $e$ and effective precessing spin $\chi_\text{p}$ are quoted at the waveform starting frequency: $5.5$~Hz for GW190521, $20$~Hz for GW200105, $23$~Hz for GW170817, and $10$~Hz otherwise.
We use \texttt{nlive}=$1000$ for all events, except GW190521, where we increase to \texttt{nlive}=$2000$ for better convergence.
The eccentricity Bayes factor $\log_{10}\mathcal{B}^{e\neq0}_{e=0}$ is computed with the Savage--Dickey density ratio~\cite{10.1214/aoms/1177693507}, with uncertainties from bootstrap resampling of the posterior; for GW200129 it is instead obtained from the nested-sampling evidences of the separate eccentric and QC runs.
The precession Bayes factor $\log_{10}\mathcal{B}^{\chi_\text{p}\neq0}_{\chi_\text{p}=0}$ is also obtained from the ratio of nested-sampling evidences of the precessing and aligned-spin analyses of each event.

Table~\ref{tab:real_events_table} reports the inferred source parameters and log Bayes factors for all the events, while Fig.~\ref{fig:ecc_chip_corner_grid} shows the corresponding joint posteriors on eccentricity $e$ and effective precessing spin $\chi_\text{p}$.
These show no appreciable correlation between $ e $ and $ \chi_\text{p} $ for any event, indicating that the two effects are not degenerate within our analyses; eccentricity correlates instead with $ \chi_\text{eff} $, mildly for most events and most strongly for GW200105, which has the tightest eccentricity measurement.
We also include in Table~\ref{tab:real_events_table} the results for GW200129 with the \texttt{gwsubtract} glitch treatment and uniform eccentricity prior for reference.
Because \texttt{SEOBNRv6EPHM} models eccentricity and precession jointly, the reported $ \log_{10}\mathcal{B}^{e\neq0}_{e=0} $ directly tests eccentricity against the QC precessing hypothesis.

Among the BBH signals other than GW200129, GW200208\_222617 shows the strongest support ($ \log_{10}\mathcal{B}^{e\neq0}_{e=0} \simeq 1.1 $, $ e \simeq 0.36 $); it is the cleanest eccentric candidate in GWTC-3 from the data-quality perspective, and likely the most robust candidate after GW200129~\cite{Romero-Shaw:2025vbc}, consistent with Refs.~\cite{Romero-Shaw:2022xko,Gupte:2024jfe, Planas:2025jny, Pompili:2026yxq}, though its false-alarm rate is relatively high~\cite{LIGOScientific:2021djp}.

The high-mass BBH events GW231223\_032836, GW190701\_203306, and GW230712\_090405 retain mild support ($ \log_{10}\mathcal{B}^{e\neq0}_{e=0} \simeq 0.6 $--$ 0.8 $) at high inferred eccentricity, in line with Refs.~\cite{Gupte:2024jfe, Gupte:2026whi, Xu:2025ajj, Pompili:2026yxq}; GW231223\_032836 and GW190701\_203306 are both affected by data-quality issues, and our analyses use glitch-subtracted frames.
GW230712\_090405, by contrast, is a clean candidate.
In aligned-spin analyses its eccentricity is favored over the QC aligned-spin hypothesis but not over the QC precessing one~\cite{Xu:2025ajj, Pompili:2026yxq}, since the signal also shows mild evidence for precession in QC analyses~\cite{LIGOScientific:2025slb}.
Modeling eccentricity and precession jointly, we instead find a mild preference for both eccentricity ($ \log_{10}\mathcal{B}^{e\neq0}_{e=0} \simeq 0.6 $) and spin precession ($ \log_{10}\mathcal{B}^{\chi_\text{p}\neq0}_{\chi_\text{p}=0} \simeq 0.7$).
GW230712\_090405 is thus a candidate for a signal from a binary that is \emph{both} eccentric and precessing, although neither effect is confidently detected, and provides an example in which the joint analysis breaks the eccentricity--spin-precession degeneracy~\cite{Romero-Shaw:2022fbf}.

The neutron-star--BH signal GW200105 shows comparable support ($ \log_{10}\mathcal{B}^{e\neq0}_{e=0} \simeq 0.7 $), consistent with Refs.~\cite{Morras:2025xfu, Planas:2025plq, Kacanja:2025kpr, Jan:2025fps, Pompili:2026yxq}, whereas GW190521 yields positive but weaker support ($ \log_{10}\mathcal{B}^{e\neq0}_{e=0} \simeq 0.4 $), in line with recent eccentric aligned-spin and unbound analyses~\cite{Ramos-Buades:2023yhy, Pompili:2026yxq, Lange:2026eqx} despite earlier eccentric and dynamical-capture interpretations~\cite{Romero-Shaw:2020thy, Gayathri:2020coq, Gamba:2021gap}.
The benchmark events GW150914 and GW170817, together with GW231123\_135430 and GW250114\_082203, disfavor eccentricity ($ \log_{10}\mathcal{B}^{e\neq0}_{e=0} < 0 $).
The results for GW250114\_082203 are consistent with the eccentric, aligned-spin analysis of Ref.~\cite{LIGOScientific:2025rid}, and with the eccentric, spin-precessing analysis with \texttt{TEOBResumS-Dal\'i} of Ref.~\cite{Chandra:2025jfc}.

GW190521 and GW231123\_135430 also show support for spin precession, consistent with QC analyses~\cite{LIGOScientific:2020iuh, LIGOScientific:2025rsn}; together with GW230712\_090405, they are the only events in our sample with such support.
GW190521 is mildly favored under both hypotheses ($ \log_{10}\mathcal{B}^{\chi_\text{p}\neq0}_{\chi_\text{p}=0} \simeq 0.9 $, $ \log_{10}\mathcal{B}^{e\neq0}_{e=0} \simeq 0.4 $), making it, with GW230712\_090405, a candidate for a binary that is eccentric \emph{and} precessing, whereas GW231123\_135430 is confidently precessing ($ \log_{10}\mathcal{B}^{\chi_\text{p}\neq0}_{\chi_\text{p}=0} \simeq 7.2 $) but disfavors eccentricity ($ \log_{10}\mathcal{B}^{e\neq0}_{e=0} \simeq -0.6 $), in agreement with Ref.~\cite{Jan:2025zcm}.
Modeling precession also sharpens the extrinsic parameters, but only where it is measured: relative to the eccentric aligned-spin analyses, the $ 90\% $ credible region in $ (d_{L}, \iota) $ narrows by more than a factor of two for GW231123\_135430 and by $ \simeq 20\% $ for GW190521 and GW230712\_090405---the three events with support for spin precession---while remaining unchanged to within a few percent for the others, consistent with the injection-recovery analyses below.
The sky localization can similarly be narrowed by up to a factor of two for the events with measured precession.

Although several of these events favor high median eccentricities ($ e \gtrsim 0.5 $), the $ 99\% $ credible region does not exclude $ e = 0 $, unlike GW200129 (see Fig.~\ref{fig:ecc_chip_corner_grid}).
Once the astrophysical prior odds between eccentric and QC mergers~\cite{Gupte:2024jfe} are folded in, as discussed above for GW200129, none of these events reaches a confident eccentric detection.

%%%%%%%%%%%%%%%%%%%%%%%%%%%
%%%%%%%%%%%%%%%%%%%%%%%%%%%

\paragraph{Injection-recovery analysis}
\phantomsection\label{sm:injections}

\begin{figure*}
\centering
\includegraphics[width=0.75\linewidth]{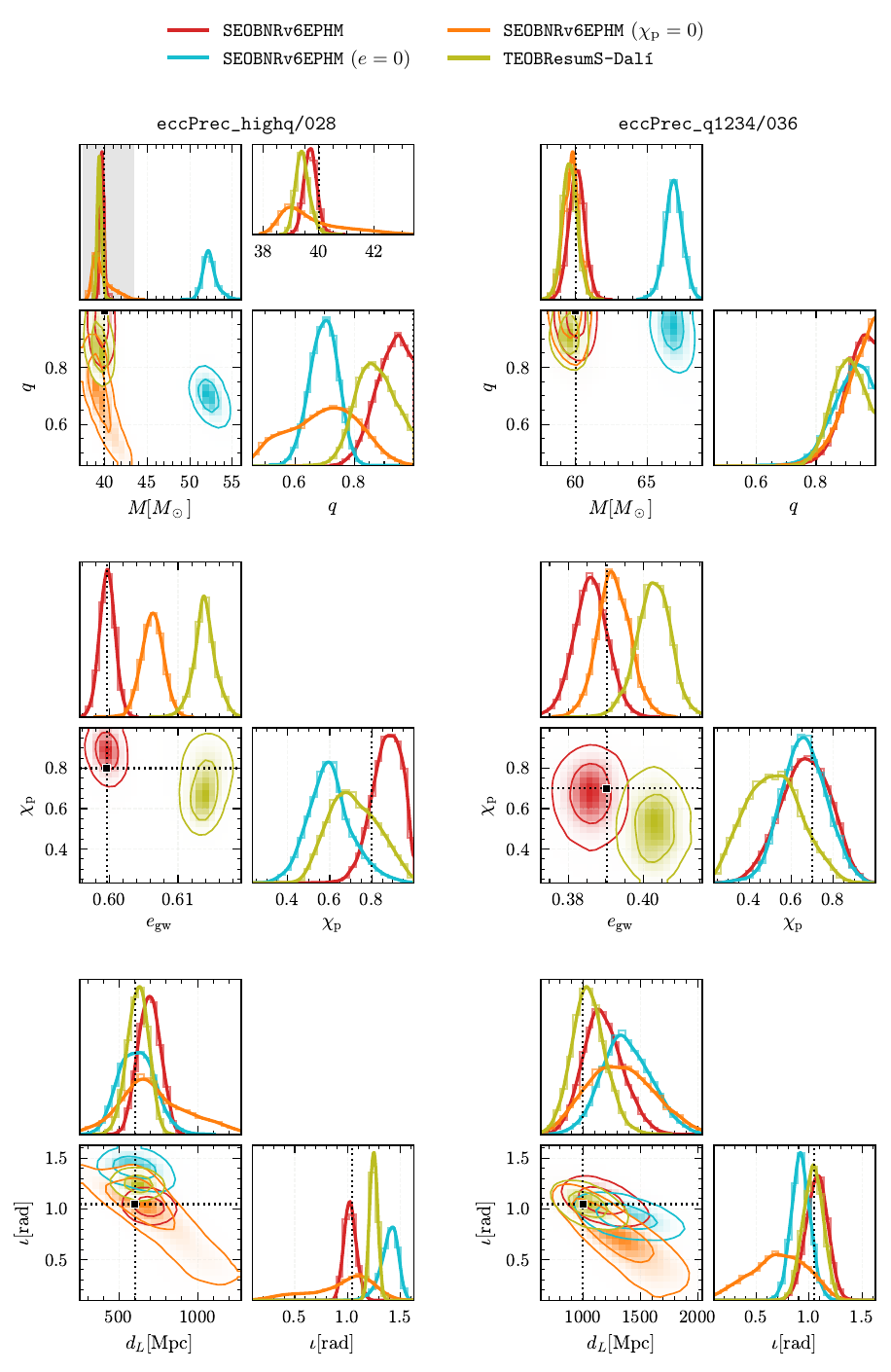}
\vspace{-5pt}
\caption{
Recovery of the two eccentric, spin-precessing synthetic NR signals (Table~\ref{tab:NR_injection_table}): \texttt{eccPrec\_highq/028} (left column) and \texttt{eccPrec\_q1234/036} (right). The rows show the masses, the eccentricity--precession pair, and the luminosity distance and inclination, for \texttt{SEOBNRv6EPHM} (red), its QC ($ e = 0 $, cyan) and aligned-spin ($ \chi_\text{p} = 0 $, orange) limits, and \texttt{TEOBResumS-Dal\'i} (yellow).
Contours represent $50\%$ and $90\%$ credible regions; one-dimensional marginals are shown both as binned histograms (step curves) and as kernel density estimates (smooth curves); dotted lines and black squares mark the injected values.
The inset in the top mass panel of the left column magnifies the shaded slice of the $M$ marginal distribution.
Spins and eccentricity are quoted at $ \langle f_\text{ref} \rangle = 10  $ Hz.
}
\label{fig:injection_corner}
\end{figure*}

To assess model accuracy in a PE context, and whether the eccentricity in GW200129 could be an artifact of waveform systematics, we inject two eccentric, spin-precessing NR signals into zero noise and recover them with \texttt{SEOBNRv6EPHM} and \texttt{TEOBResumS-Dal\'i}, following the setup of Ref.~\cite{Pompili:2026yxq}.
We use the private \texttt{SXS} simulations \texttt{eccPrec\_highq/028} and \texttt{eccPrec\_q1234/036} with parameters listed in Table~\ref{tab:NR_injection_table}, and we present main results in Fig.~\ref{fig:injection_corner}.
A broader injection campaign is left for future work.

\begin{table}
\caption{
Parameters of the two eccentric, spin-precessing NR injections, with $ \chi_\text{eff} $, $ \chi_\text{p} $ and $ e_\text{gw} $ quoted at $ \langle f_\text{ref} \rangle= 10 $ Hz.
}
\input{tab/NR_injection_table}
\label{tab:NR_injection_table}
\end{table}

Both injected NR waveforms are from equal-mass BBHs with strongly misaligned spins, and they probe two different regimes: \texttt{028} has $ |\bm{\chi}_1| = |\bm{\chi}_2| = 0.80 $ with the primary spin almost fully in the orbital plane, so that $ \chi_\text{p} = 0.80 $ and $ \chi_\text{eff} < 0 $, whereas \texttt{036} has $ |\bm{\chi}_1| = 0.80 $, $ |\bm{\chi}_2| = 0.30 $ at moderate tilts, giving $ \chi_\text{p} = 0.70 $ and $ \chi_\text{eff} > 0 $. 
Their GW eccentricities are $ e_\text{gw} \simeq 0.60 $ and $ 0.39 $ at a reference frequency of $ 10 $ Hz, respectively.
The simulation \texttt{eccPrec\_q1234/036} has parameter values (total mass, mass ratio, and eccentricity) close to those inferred for GW200129, while \texttt{eccPrec\_highq/028} is more extreme in spin and eccentricity, and longer in duration, so that it is more challenging to recover.
\texttt{SEOBNRv6EPHM} reproduces the two NR waveforms with mismatches of $\sim 1.6$\% for \texttt{028} and 0.4\% for \texttt{036}, comparable to the model's median mismatch against the NR simulations considered; the two injections are therefore representative of its typical accuracy.

The total mass of each injection is chosen so that the orbit-averaged $ (2,2) $ frequency at the NR reference epoch is close to $ 10 $ Hz ($ M = 40 $ and $ 60 \, \solarmass $), ensuring that the full length of the available NR waveform is employed.
The recovery waveform uses a corresponding starting frequency of $ 10 $ Hz.
The eccentricity and the spins are also quoted at a common reference frequency $ \langle f_\text{ref} \rangle = 10 $ Hz, with the eccentricity measured with \texttt{gw\_eccentricity}~\cite{Shaikh:2023ypz,Shaikh:2025tae}.
Measuring $ e_\text{gw} $ at $ 10 $ Hz requires the waveform to extend below it, which \texttt{SEOBNRv6EPHM} provides through its backward-in-time evolution but \texttt{TEOBResumS-Dal\'i} does not; the NR $ e_\text{gw} $ curve is likewise only measurable above $ \simeq 10.5 $ Hz for these two simulations at the chosen total masses.
For NR and \texttt{TEOBResumS-Dal\'i}, we therefore sample $ e_\text{gw} $ on a grid of frequencies above $ 10 $ Hz and extrapolate down to it (to stabilize the extrapolation, we divide out the prediction of the Peters relation~\cite{Peters:1964zz} and fit the slowly drifting residual).
We validate this procedure against \texttt{SEOBNRv6EPHM}, where the $ 10 $ Hz value is directly measurable: the extrapolation is accurate to $ \lesssim 10^{-4} $ in $ e_\text{gw} $, far below the statistical uncertainties.
The injections use the same settings as the GW200129 analyses, with an \texttt{A+} PSD in a H1--L1 network, and reach network SNRs of $ 42 $ and $ 38 $, both above the $ 25.6 $ value recovered for GW200129.

The top two rows of Fig.~\ref{fig:injection_corner} show the recovery of the \emph{intrinsic} parameters against the injected values, for \texttt{SEOBNRv6EPHM} and its two variants ($ e = 0 $ and $ \chi_\text{p} = 0 $), and the independent eccentric, spin-precessing model \texttt{TEOBResumS-Dal\'i}.
\texttt{SEOBNRv6EPHM} is the only model that recovers every intrinsic parameter without significant bias, matching $ e_\text{gw} $ to within $ 0.004 $ of the injected value and every other parameter to better than one half-width of its $ 90\% $ credible interval.
\texttt{TEOBResumS-Dal\'i}, although also eccentric and spin-precessing, overestimates $ e_\text{gw} $ by $ \simeq 0.013 $ (a $ 2 $--$ 5 $ half-width bias), mildly favors unequal masses (injected $ q = 1 $) and, for \texttt{036}, slightly underestimates $ \chi_\text{p} $; the other parameters are recovered comparably well.
Neglecting either effect biases the recovery: the QC limit ($ e = 0 $) reabsorbs the missing eccentricity by shifting the recovered masses, overestimating the total mass by $ \simeq 12 $ and $ 7 \, \solarmass $ for \texttt{028} and \texttt{036}, about seven half-widths in both cases; the aligned-spin limit ($ \chi_\text{p} = 0 $) recovers the eccentricity similarly well, but a mass ratio that excludes the injected $ q = 1 $ at 90\% credibility for \texttt{028}.
Both variants also leave signal power unrecovered.
For \texttt{028}, the recovered network matched-filter SNR is $\simeq 42.1 $ for the full model and $ 41.8 $ for \texttt{TEOBResumS-Dal\'i}, but falls to $ 30.5 $ ($ e = 0 $) and $ 38.1 $ ($ \chi_\text{p} = 0 $); for \texttt{036} the corresponding values are $\simeq 38.0 $, $ 37.9 $, $ 35.2 $ and $ 37.1 $, respectively.
That the recovery is unbiased at an SNR higher than GW200129's supports the eccentric interpretation of that event against waveform systematics.

The bottom row of Fig.~\ref{fig:injection_corner} shows the luminosity distance $ d_{L} $ and inclination $ \iota $, which set the signal amplitude. Since eccentricity and precession also modulate the amplitude, neglecting either effect can be reabsorbed into biases of $ d_{L} $ and $ \iota $.
The injected values lie inside the $ 90\% $ credible region of \texttt{SEOBNRv6EPHM} for both injections.
For \texttt{028} they fall instead outside the $ 90\% $ region of both the QC limit and \texttt{TEOBResumS-Dal\'i}.
The aligned-spin limit does contain the injected values, but recovers the parameters less precisely: its $ 90\% $ area is $ 410\,\mathrm{Mpc\,rad} $, against $ 57\,\mathrm{Mpc\,rad} $ for the full model.
The effect is milder for \texttt{036}: there every model contains the injected values within the $90\%$ credible region;
the two eccentric, spin-precessing models nonetheless recover the parameters more precisely than either variant.

By capturing eccentricity and precession together, \texttt{SEOBNRv6EPHM} breaks the distance--inclination degeneracy, recovering $ d_{L} $ and $ \iota $ both accurately and precisely.
Such unbiased, precise measurements are crucial across several applications:
in multimessenger astrophysics, $ d_{L} $ and $ \iota $ help localize electromagnetic counterparts and constrain their geometry (e.g., jet viewing angles); in cosmology, $ d_{L} $ is the key standard-siren observable for the Hubble constant, so a bias in $ d_{L} $ propagates to the inferred expansion rate; and, because source-frame masses are obtained from $ d_{L} $ through the redshift, sharper distances would yield sharper masses, benefiting population analyses.

%%%%%%%%%%%%%%%%%%%%%%%%%%%
%%%%%%%%%%%%%%%%%%%%%%%%%%%

\paragraph{Additional benchmark results}
\phantomsection\label{sm:benchmarks}

\begin{figure}
\hspace{-5pt}
\includegraphics[width=\linewidth]{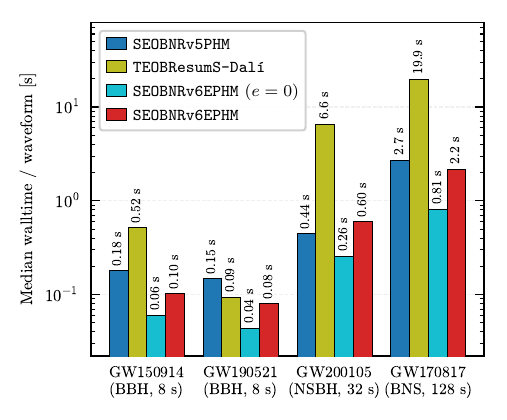}
\vspace{-5pt}
\caption{
Median walltime per waveform evaluation for the four GW events of Table~\ref{tab:benchmarks}, using the spin-precessing models \texttt{SEOBNRv5PHM} (QC), \texttt{SEOBNRv6EPHM} in its QC ($ e = 0 $) and eccentric configurations, and \texttt{TEOBResumS-Dal\'i} (eccentric).
For each event, $ 128 $ samples are drawn from the \texttt{SEOBNRv6EPHM} posterior, and the corresponding waveforms are generated with every model under identical settings.
The binary type and segment duration are indicated below each group of bars; values are printed above the bars.
}
\label{fig:timing_comparison}
\end{figure}
\begin{figure}
\hspace{-2pt}
\vspace{6pt}
\includegraphics[width=0.995\linewidth]{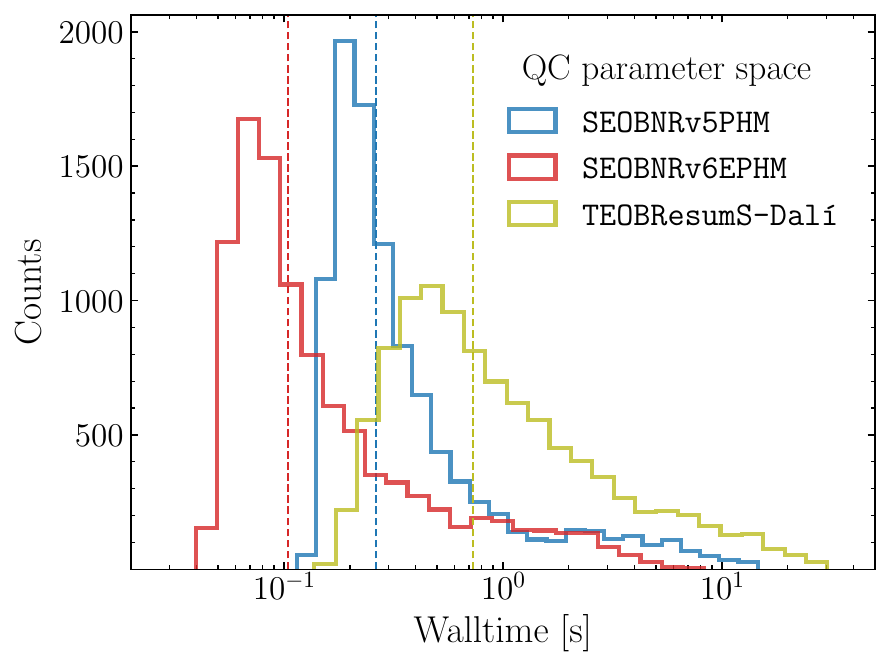}
\includegraphics[width=\linewidth]{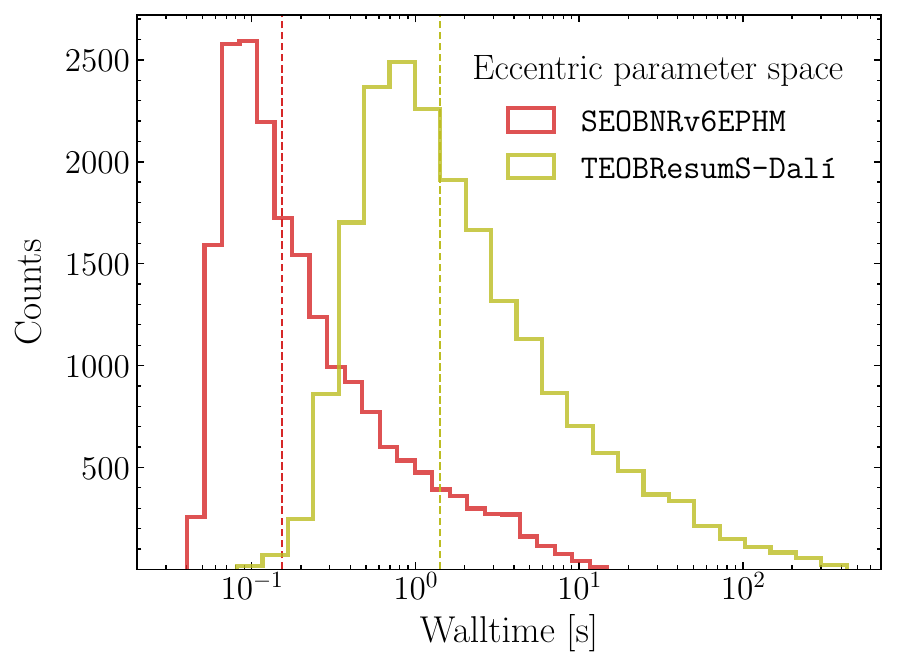}
\vspace{-10pt}
\caption{
Distributions of the waveform-evaluation walltime for each model, over randomly drawn QC (top) and eccentric (bottom) configurations, with waveforms generated from an orbit-averaged frequency $ \langle f _{ \text{start}} \rangle = 10 \, $Hz at a sampling rate of $ 2048 \, $Hz.
The QC set comprises $ 10^4 $ configurations with $ q \in [1, 10] $, $ M \in [5, 100] \, \solarmass $, and spins $ \bm{\chi}_i $ drawn uniformly from the unit ball; the eccentric set comprises $ 2 \times 10^4 $ configurations that additionally sample $ e \in [0, 0.5] $ with $ \zeta = 0 $.
Dashed lines mark the median of each distribution.
}
\label{fig:benchmarks}
\end{figure}

We complement the PE benchmarks of the main text with measurements of the waveform-evaluation walltimes for \texttt{SEOBNRv5PHM}, \texttt{SEOBNRv6EPHM}, and \texttt{TEOBResumS-Dal\'i}.

Since the higher computational cost of \texttt{TEOBResumS-Dal\'i} would make PE analyses of the longer signals infeasible, we compare all models on a common footing: for each of the four benchmark events in Table~\ref{tab:benchmarks}, we draw $ 128 $ samples from the \texttt{SEOBNRv6EPHM} posteriors, and generate waveforms with \texttt{SEOBNRv5PHM}, \texttt{SEOBNRv6EPHM} in its QC limit ($ e = 0 $), and the full eccentric, spin-precessing versions of \texttt{SEOBNRv6EPHM} and \texttt{TEOBResumS-Dal\'i}.
Every model uses the same parameters and settings as the corresponding analysis (segment duration, sampling rate, frequency range, and reference frequency).
Each evaluation includes the conditioning of the time-domain waveform and its transformation to the frequency domain, as in the PE analyses.
All timings are performed on a single core of an AMD EPYC 7351 processor.

Figure~\ref{fig:timing_comparison} shows the resulting median evaluation walltimes.
In its QC limit, \texttt{SEOBNRv6EPHM} is $ \sim 2$--$ 3 $ times faster than \texttt{SEOBNRv5PHM}, while the full eccentric, spin-precessing model is comparable in cost.
Against \texttt{TEOBResumS-Dal\'i}, \texttt{SEOBNRv6EPHM} is comparable for the short, high-mass signal GW190521, $ \sim 5 $ times faster for GW150914, and an order of magnitude faster for the neutron-star--BH and binary neutron star signals GW200105 and GW170817---a trend with signal duration that makes eccentric, spin-precessing analyses of long signals feasible.
For the \texttt{SEOBNR} models, these per-waveform ratios are close to the corresponding ratios of PE runtimes in Table~\ref{tab:benchmarks}, confirming that the runtime differences between the PE runs are mostly determined by the model cost.

Beyond the benchmark events, Fig.~\ref{fig:benchmarks} shows the walltime distributions for generating the time-domain polarizations of randomly drawn QC and eccentric spin-precessing configurations, sampled over ranges that overlap with the observed BBH population~\cite{LIGOScientific:2026ctl} (for eccentric configurations, we set $ \zeta = 0 $, since the radial anomaly does not significantly change the waveform length).
For the QC parameter space (top panel of Fig.~\ref{fig:benchmarks}), \texttt{SEOBNRv6EPHM} is faster than \texttt{SEOBNRv5PHM} and \texttt{TEOBResumS-Dal\'i} by median factors of $ 2 $ and $ 6 $, respectively.
For the eccentric case (bottom panel), \texttt{SEOBNRv6EPHM} is faster than \texttt{TEOBResumS-Dal\'i} by a median factor of $ \sim 10 $, with the slower tails of the distributions (corresponding to the longest systems, with high mass ratios or small total masses) separated by a factor of $ \sim 30 $.
These findings are consistent with, and extend, the per-event results of Table~\ref{tab:benchmarks} and Fig.~\ref{fig:timing_comparison}.

Therefore, \texttt{SEOBNRv6EPHM} is the fastest model across the sampled eccentric, spin-precessing parameter space, and its computational efficiency enables large-scale analyses of fully generic BBH configurations.

%%%%%%%%%%%%%%%%%%%%%%%%%%%
%%%%%%%%%%%%%%%%%%%%%%%%%%%
%%%%%%%%%%%%%%%%%%%%%%%%%%%

%\clearpage

%\twocolumngrid

\bibliography{../references/references}

%To compile the references, one must compile the BibTex on the TexShop menu, or use the command cmd+shift+b, everytime there is a change in the references.
%To autocomplete a reference, start writing \cite{
%and then the letter of the autor and press F5

%To autogenerate a citation key in bibdesk, use Command+K

% To create a simlink or alias, go to the directory in which the tex file lives, and type something like:
% ln -s /Users/a.gamboa/Documents/manuscripts/references/references.bib  references.bib

% Sort references with
% btidy references.bib

\end{document}

%% file: tab/benchmarks.tex
\renewcommand{\arraystretch}{1.25}
\setlength{\tabcolsep}{1pt}
\small

\begin{ruledtabular}
\begin{tabular*}{\columnwidth}{l@{\extracolsep{\fill}} cc cc cc}
 & \multicolumn{2}{c}{\texttt{v5PHM}}
 & \multicolumn{2}{c}{\texttt{v6EPHM} ($e{=}0$)}
 & \multicolumn{2}{c}{\texttt{v6EPHM}}
\\
Event
 & Runtime & $\mathcal{L}$ eval.
 & Runtime & $\mathcal{L}$ eval.
 & Runtime & $\mathcal{L}$ eval.
\\
\colrule
GW150914 & 1d 15h & 4.9 & 13h & 4.9 & 1d 1h & 5.2 \\
GW190521 & 1d 6h & 4.7 & 9h & 4.6 & 1d 1h & 6.9 \\
GW200105 & 1d 4h & 5.4 & 16h & 5.4 & 1d 17h & 7.0 \\
GW170817 & 4d 18h & 4.9 & 2d 2h & 5.1 & 5d 5h & 5.6 \\

\end{tabular*}
\end{ruledtabular}

%% file: tab/GW200129_table.tex
\renewcommand{\arraystretch}{1.25}
\setlength{\tabcolsep}{1pt}
\begin{tabular*}{\textwidth}{l@{\extracolsep{\fill}}lcccccc}
\toprule
 & & \multicolumn{3}{c}{$\log_{10}\mathcal{B}$ against} & & & \\
\cmidrule(lr){3-5}
Glitch subtraction & $e$ prior & \texttt{v6EPHM} ($e = 0$) & \texttt{NRSur} ($e = 0$) & \texttt{v6EPHM} ($\chi_\mathrm{p} = 0$) & $e_{10\,\mathrm{Hz}}$ & $e_{\mathrm{gw},10\,\mathrm{Hz}}$ & $\chi_\mathrm{p}$ \\
\midrule
\multirow{2}{*}{No mitigation} & Uniform & \cellcolor[rgb]{0.490,0.808,0.627} $+5.66$ & \cellcolor[rgb]{0.512,0.814,0.643} $+5.38$ & \cellcolor[rgb]{0.908,0.936,0.920} $+0.49$ & $0.29^{+0.06}_{-0.06}$ & $0.29^{+0.06}_{-0.06}$ & $0.32^{+0.30}_{-0.21}$ \\
 & Log-uniform & \cellcolor[rgb]{0.555,0.828,0.673} $+4.85$ & \cellcolor[rgb]{0.577,0.834,0.688} $+4.57$ & \cellcolor[rgb]{0.910,0.784,0.772} $-0.32$ & $0.28^{+0.04}_{-0.05}$ & $0.28^{+0.04}_{-0.05}$ & $0.33^{+0.35}_{-0.21}$ \\
\multirow{2}{*}{gwsubtract} & Uniform & \cellcolor[rgb]{0.573,0.833,0.685} $+4.64$ & \cellcolor[rgb]{0.595,0.840,0.701} $+4.35$ & \cellcolor[rgb]{0.933,0.944,0.938} $+0.18$ & $0.26^{+0.05}_{-0.07}$ & $0.26^{+0.05}_{-0.07}$ & $0.27^{+0.31}_{-0.19}$ \\
 & Log-uniform & \cellcolor[rgb]{0.631,0.851,0.726} $+3.92$ & \cellcolor[rgb]{0.652,0.858,0.741} $+3.63$ & \cellcolor[rgb]{0.885,0.677,0.656} $-0.54$ & $0.25^{+0.04}_{-0.07}$ & $0.25^{+0.04}_{-0.07}$ & $0.29^{+0.32}_{-0.21}$ \\
\multirow{2}{*}{NLSUB} & Uniform & \cellcolor[rgb]{0.656,0.859,0.743} $+3.59$ & \cellcolor[rgb]{0.663,0.861,0.749} $+3.51$ & \cellcolor[rgb]{0.926,0.853,0.846} $-0.19$ & $0.24^{+0.05}_{-0.07}$ & $0.24^{+0.05}_{-0.07}$ & $0.27^{+0.29}_{-0.19}$ \\
 & Log-uniform & \cellcolor[rgb]{0.710,0.875,0.781} $+2.96$ & \cellcolor[rgb]{0.717,0.878,0.786} $+2.87$ & \cellcolor[rgb]{0.851,0.533,0.502} $-0.82$ & $0.23^{+0.06}_{-0.07}$ & $0.23^{+0.06}_{-0.07}$ & $0.26^{+0.28}_{-0.18}$ \\
\multirow{2}{*}{BayesWave A} & Uniform & \cellcolor[rgb]{0.854,0.920,0.882} $+1.19$ & \cellcolor[rgb]{0.843,0.916,0.875} $+1.30$ & \cellcolor[rgb]{0.914,0.801,0.789} $-0.29$ & $0.18^{+0.07}_{-0.08}$ & $0.17^{+0.07}_{-0.08}$ & $0.37^{+0.35}_{-0.23}$ \\
 & Log-uniform & \cellcolor[rgb]{0.890,0.931,0.907} $+0.73$ & \cellcolor[rgb]{0.882,0.929,0.902} $+0.84$ & \cellcolor[rgb]{0.859,0.566,0.537} $-0.76$ & $0.13^{+0.08}_{-0.13}$ & $0.13^{+0.08}_{-0.13}$ & $0.47^{+0.33}_{-0.29}$ \\
\multirow{2}{*}{BayesWave B} & Uniform & \cellcolor[rgb]{0.753,0.889,0.812} $+2.43$ & \cellcolor[rgb]{0.753,0.889,0.812} $+2.43$ & \cellcolor[rgb]{0.936,0.895,0.891} $-0.11$ & $0.22^{+0.05}_{-0.07}$ & $0.22^{+0.05}_{-0.07}$ & $0.34^{+0.36}_{-0.21}$ \\
 & Log-uniform & \cellcolor[rgb]{0.807,0.905,0.849} $+1.77$ & \cellcolor[rgb]{0.807,0.905,0.849} $+1.77$ & \cellcolor[rgb]{0.856,0.556,0.527} $-0.77$ & $0.20^{+0.06}_{-0.09}$ & $0.20^{+0.06}_{-0.09}$ & $0.35^{+0.33}_{-0.22}$ \\
\multirow{2}{*}{BayesWave C} & Uniform & \cellcolor[rgb]{0.731,0.882,0.796} $+2.66$ & \cellcolor[rgb]{0.756,0.890,0.814} $+2.39$ & \cellcolor[rgb]{0.911,0.937,0.923} $+0.46$ & $0.23^{+0.05}_{-0.08}$ & $0.23^{+0.05}_{-0.08}$ & $0.41^{+0.34}_{-0.25}$ \\
 & Log-uniform & \cellcolor[rgb]{0.785,0.899,0.834} $+1.99$ & \cellcolor[rgb]{0.810,0.906,0.852} $+1.71$ & \cellcolor[rgb]{0.924,0.843,0.835} $-0.21$ & $0.22^{+0.05}_{-0.09}$ & $0.22^{+0.05}_{-0.09}$ & $0.41^{+0.34}_{-0.24}$ \\
\bottomrule
\end{tabular*}

%% file: tab/real_events_table.tex
\renewcommand{\arraystretch}{1.5}
\begin{tabular*}{\textwidth}{c@{\extracolsep{\fill}} c c c c c c c}

\hline
\hline

Event
& $M_{\rm src}/\solarmass$
& $1/q$
& $\chi_{\text{eff}}$
& $\chi_{\text{p}}$
& $e$
& $\log_{10} \mathcal{B}^{e \neq 0}_{e = 0}$
& $\log_{10} \mathcal{B}^{\chi_\text{p}\neq0}_{\chi_\text{p}=0}$
\\ [0.05cm]

\hline

GW200129\textsuperscript{\ddag}
    & $60.5^{+3.4}_{-2.9}$ & $0.72^{+0.19}_{-0.15}$ & $0.01^{+0.09}_{-0.08}$ & $0.27^{+0.31}_{-0.19}$ & $0.26^{+0.05}_{-0.07}$ & \cellcolor[rgb]{0.490,0.808,0.627} $4.64^{+0.10}_{-0.10}$ & \cellcolor[rgb]{0.936,0.945,0.940} $0.18$
    \\

\hline

GW200208\_222617
    & $42.0^{+10.5}_{-6.5}$ & $0.57^{+0.36}_{-0.28}$ & $0.08^{+0.29}_{-0.21}$ & $0.42^{+0.43}_{-0.30}$ & $0.36^{+0.11}_{-0.15}$ & \cellcolor[rgb]{0.839,0.915,0.872} $1.11^{+0.25}_{-0.12}$ & \cellcolor[rgb]{0.944,0.947,0.945} $0.07$
    \\

\hline

GW231223\_032836
    & $77.6^{+18.3}_{-13.3}$ & $0.74^{+0.23}_{-0.34}$ & $0.03^{+0.31}_{-0.37}$ & $0.45^{+0.41}_{-0.33}$ & $0.46^{+0.08}_{-0.28}$ & \cellcolor[rgb]{0.875,0.926,0.897} $0.75^{+0.13}_{-0.08}$ & \cellcolor[rgb]{0.942,0.921,0.919} $-0.03$
    \\

\hline

GW200105\_162426
    & $10.5^{+0.7}_{-0.9}$ & $0.23^{+0.06}_{-0.03}$ & $-0.08^{+0.10}_{-0.15}$ & $0.11^{+0.11}_{-0.08}$ & $0.11^{+0.03}_{-0.02}$ & \cellcolor[rgb]{0.879,0.927,0.900} $0.72^{+0.18}_{-0.15}$ & \cellcolor[rgb]{0.851,0.533,0.502} $-0.45$
    \\

\hline

GW190701\_203306
    & $96.3^{+12.6}_{-10.8}$ & $0.76^{+0.22}_{-0.33}$ & $-0.07^{+0.25}_{-0.29}$ & $0.43^{+0.42}_{-0.32}$ & $0.54^{+0.08}_{-0.35}$ & \cellcolor[rgb]{0.879,0.927,0.900} $0.72^{+0.19}_{-0.09}$ & \cellcolor[rgb]{0.944,0.947,0.945} $0.06$
    \\

\hline

GW230712\_090405
    & $61.8^{+17.9}_{-17.2}$ & $0.44^{+0.42}_{-0.17}$ & $0.20^{+0.38}_{-0.48}$ & $0.58^{+0.32}_{-0.38}$ & $0.63^{+0.07}_{-0.47}$ & \cellcolor[rgb]{0.890,0.931,0.907} $0.59^{+0.06}_{-0.06}$ & \cellcolor[rgb]{0.908,0.936,0.920} $0.67$
    \\

\hline

GW190521
    & $163.5^{+23.1}_{-18.3}$ & $0.69^{+0.26}_{-0.27}$ & $-0.02^{+0.35}_{-0.28}$ & $0.74^{+0.21}_{-0.41}$ & $0.63^{+0.04}_{-0.53}$ & \cellcolor[rgb]{0.911,0.937,0.923} $0.38^{+0.05}_{-0.04}$ & \cellcolor[rgb]{0.890,0.931,0.907} $0.91$
    \\

\hline

GW231123\_135430
    & $237.7^{+27.8}_{-25.4}$ & $0.82^{+0.16}_{-0.21}$ & $0.42^{+0.20}_{-0.22}$ & $0.76^{+0.16}_{-0.22}$ & $0.10^{+0.18}_{-0.09}$ & \cellcolor[rgb]{0.910,0.784,0.772} $-0.60^{+0.03}_{-0.02}$ & \cellcolor[rgb]{0.490,0.808,0.627} $7.25$
    \\

\hline

GW150914
    & $63.6^{+3.7}_{-3.0}$ & $0.86^{+0.13}_{-0.21}$ & $-0.08^{+0.14}_{-0.13}$ & $0.40^{+0.39}_{-0.30}$ & $0.06^{+0.08}_{-0.06}$ & \cellcolor[rgb]{0.898,0.732,0.716} $-0.78^{+0.04}_{-0.03}$ & \cellcolor[rgb]{0.936,0.945,0.940} $0.21$
    \\

\hline

GW170817
    & $2.76^{+0.05}_{-0.03}$ & $0.75^{+0.20}_{-0.12}$ & $0.00^{+0.02}_{-0.02}$ & $0.02^{+0.02}_{-0.02}$ & $0.008^{+0.006}_{-0.007}$ & \cellcolor[rgb]{0.863,0.582,0.555} $-1.33^{+0.05}_{-0.07}$ & \cellcolor[rgb]{0.929,0.863,0.856} $-0.09$
    \\

\hline

GW250114\_082203
    & $65.9^{+1.1}_{-1.1}$ & $0.96^{+0.04}_{-0.06}$ & $-0.03^{+0.04}_{-0.04}$ & $0.10^{+0.19}_{-0.07}$ & $0.012^{+0.021}_{-0.011}$ & \cellcolor[rgb]{0.851,0.533,0.502} $-1.51^{+0.02}_{-0.02}$ & \cellcolor[rgb]{0.894,0.716,0.698} $-0.25$
    \\

\hline

\hline

\end{tabular*}

%% file: tab/NR_injection_table.tex
\renewcommand{\arraystretch}{1.15}
\begin{tabular*}{\columnwidth}{l@{\extracolsep{\fill}} cc}
\hline\hline
 & \texttt{eccPrec\_highq/028} & \texttt{eccPrec\_q1234/036} \\
\hline
$M / \solarmass$ & $40$ & $60$ \\
$q$ & $1.00$ & $1.00$ \\
$d_{L} / \text{Mpc}$ & $600$ & $1000$ \\
$\iota$ [deg] & $60$ & $60$ \\
$\chi_{\rm eff}$ & $-0.20$ & $+0.28$ \\
$\chi_{\rm p}$ & $0.80$ & $0.70$ \\
$e_{\rm gw}$ & $0.60$ & $0.39$ \\
Network SNR & $42.1$ & $38.0$ \\
\hline\hline
\end{tabular*}